\documentclass[12pt]{report}
\usepackage{graphics}
\usepackage{epsfig}

\begin{document}
\thispagestyle{empty}
\pagenumbering{roman}
\begin{center}
{\textbf{ Diffraction of Electromagnetic Wave by Circular Disk and Circular Hole}}
\vfill
\bigskip
\begin{figure}[h]
\begin{center}
\includegraphics[width=6cm, height=4.5cm]{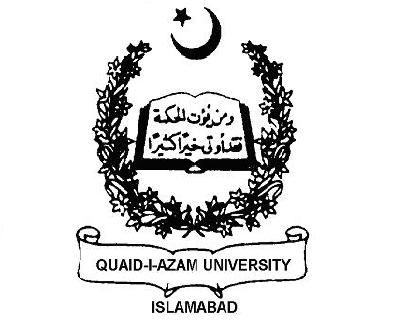}
\end{center}
\end{figure}
\vfill

Muhammad Adnan Shahzad
\vfill

\textbf{DEPARTMENT OF ELECTRONICS}\\
\textbf{QUAID-I-AZAM UNIVERSITY}\\
\textbf{ISLAMABAD, PAKISTAN}
\vfill
2006
\end{center}

\pagebreak
\thispagestyle{empty}
@finalout
\begin{center}
{\textbf{ DIFFRACTION OF ELECTROMAGNETIC WAVE BY CIRCULAR DISK AND CIRCULAR HOLE}}
\vfill
Thesis by\\
Muhammad Adnan Shahzad
\vfill
In Partial Fulfillment of the Requirements \\
for the Degree of\\
Master of Philosophy
\vfill
\textbf{Department of Electronics}\\
\textbf{Quaid-I-Azam University}\\
\textbf{Islamabad, Pakistan}\\
\vfill
2006
\end{center}
\pagebreak 

\thispagestyle{empty}
\begin{center}
\vfill
{\textbf{CERTIFICATE}}
\end{center}
 \paragraph{} Certified that the work contained in this desseration was carried out by Muhammad Adnan Shahzad under my
 supervision.\vspace{12pt}
 \begin{flushright}
 (\textbf{Dr. Qaisar Abbas Naqvi})\\
  Associate Professor\\
  Department  of Electronics,\\
  Quaid-I-Azam University
 Islamabad, \\Pakistan
 \end{flushright}
 \vspace{2.0cm}
 \begin{flushleft}
 Submitted Through\\
 \vspace{1.0cm}
 (\textbf{Prof. Dr. Azhar Abbas Rizvi})\\
 Chairman\\
 Department of Electronics,\\
 Quaid-I-Azam University Islamabad,\\
 Pakistan
 \end{flushleft}
 \pagebreak
\thispagestyle{empty}
\begin{center}
\vfill
\textbf{DEDICATION}\\
\bigskip
This dessertation is dedicated to my brothers, \\
Saqib  and Sajjawal 
\end{center}
\pagebreak

\thispagestyle{empty}
\begin{center}
\vfill
\vfill
{\textbf{ACKNOWLEDGMENT}}
\end{center}
 \paragraph{} All praise and thanks are due to Allah Who is the Lord and Sustainer of the universe.\vspace{12pt}
\smallskip
\indent 
I would like to thanks from the depth of my heart to Dr. Qaisar Abbas Naqvi who encouraged me a lot and supervised me with great patience and co-operation. This work could not have completed without his valuable guidance, encouragement and constructive criticisms. Thnaks to the Chairman Deparment of Electronics Prof. Dr. A.A Rizvi for granting me research facilities.\\
\smallskip
\indent 
I wish to thanks Prof. Kohei Hongo, Toho University Japan, who provided me literature and help me in  computational work during my research. \\
\smallskip
\indent
 I wish to acknowledge discussions with my lab-fellows, Muhammad Naveed, Shakeel Ahmed and Abdul Ghafar on topics related and unrelated to electromagnetic field theory. Thnaks to Fazli Manan and Husnul Maab whose guided me at each stage and also assisted me in my research activity.\\ 
\smallskip
\indent I am gratful to my parents who gave me a long leave of absence from my responsiblities. I have no words to thank my father for his encouragement, financial assistance and guidance during my formative years. I am also gratful to my brothers and sisters for their encouragment, love and warm wishes.
\begin{flushright}
Muhammad A. Shahzad
\end{flushright}
\pagebreak

\thispagestyle{empty}
\begin{center}
\vfill
{\textbf{ABSTRACT}}

\end{center}
\paragraph{}The problems of diffraction of an electromagnetic plane wave by a perfectly conducting circular disk and its complementary problem, diffraction by a circular hole in an infinite conducting plate, are rigorously solved using the method of the kobayashi petential. Th mathematical formulation involves dual integral equations derived from the potential integral and boundary condition on the plane where a disk or hole is located. The field is expressed by a linear combination of function which satisfy the required boundary conditions except on the disk or hole. It may also be varified that the solution for the disk and the hole satisfy Babinet's principal. The weighting function in the potential integrals are determined by applying the properties of the Weber-Schafheitlin's integral and the soluation is obtained in the form of a matrix equation. Matrix elements of the equations for the expansion coefficients are given by three kinds of infinite integrals and the series soluation for these infinite integral are derived. For the varification of these series soluation, the numerical integral are also derived and the results are computed numerically for conformation, which is fairly good. Illustrative computations are given for the far diffracted field pattern. The results of the far-field patterns are compared with the results obtained from physical optics (PO). The agreement is fairly good.

\newpage
\renewcommand{\contentsname}{Table of Contents}

\tableofcontents
\cleardoublepage
\pagebreak
\pagenumbering{arabic}

\chapter{Introduction}\label{Introduction}
\paragraph{}Over several decades, electromagnetic (EM) scattering from a circular disk has attracted researchers with a strong intrest in the (monostatic or backscattering) radar cross section (RCS's) of large dynamic range.Usually, as reviewed by Duan and Rahmat-Samii [1], the following groups of techniques are applied to the analysis of the electromagnetic scattering from circular disk.
\indent The first type is the physical optics (PO) [2] which is an approximate technique and is accurate for predicting the far-filed pattern near the main beam. The second type is the physical theory of diffraction (PTD) [3] which is more accurate than the PO technique since the equivalent edge current is applied and the caustic singularities in the original ray tracing are eliminated. This method is further modified (a) by Ando [4] using equivalent edge currents, (b) by Mitzner [5] utilizing the incremental length diffraction coefficients, and (c) by Michaeli [6,7] using surface-to-edge integral and the fringe current radiation integral over the ray coordinates instead of the normal coordinates. The third type is the geometrical theory of diffraction (GTD) [9-13], which has similar accuracy to the PTD. This method was also modified into (a) uniform geometrical theory of diffraction by Kouyoumijian and Pathak [14], (b) uniform asymptotic technique by Ahluwalia et al. [15] and Lee and Deschamps [16], and (c) High-Order Geometrical Theory of Diffraction by Bechtel [17] and Ryan and Peters [18] (of the Ist order), by Knott et al. [19] (of the second order), and  by Marsland et al. [20] (of the higher-order). The fourth type is the method of moments (MoM) or moment method (MM) [1] that is considered to be numerically exact. The Hybrid Asymptotic Moment Method is implemented by Kim and Thiele first who found the induced currents on the scatterer surface. This method was further modified by Kaye, Murthy and Thiele [21,22], and thus the fifth type of methods was formed i.e., the hybrid-iterative method which employs the magnetic field integral equation for the induced currents to solve the scattering problem.\vspace{12pt}
\indent The diffraction of electromagnetic waves by a circular disk and a circular hole is solved by Bouwkamp [23] by using the power series expansion, by Meixner and Anderjewski [24], by Andrejewski [25] and by Flammer [26] by using the spheroidal wave function, and by Levine and Schwinger [27] by using the variation method. The variation method is not sufficient since it depends on the trial functions. The spheroidal wave function is not adequate in extending it to a complicated problem, but give great numerical results for the case of circular disk. Following the idea of Meixner and Anderjewski, Nomura and Katsura [28] reformulated this problem by using the Weber-Schafheitlin's discontinuous integral (Kobayashi Potential). Here we discuss an alternative method to that by Nomura and Katsura.\vspace{12pt}
\indent Most of the researcher have used an integral equation for unknown equivalent surface current density on the aperture or disk. This integral equation is reduce to matrix equation via the method of moments (MoM). In this dissertation, rigorous solution to the problem of a plane wave scattering by a circular conducting disk and its complementary problem, diffraction by a circular hole on a perfectly conducting plane, are derived using the method of the Kobayashi potential method [29], [30]. This method has been applied to various kinds of problem such as the potential problems of electrified circular disks [31],[32], the diffraction of acoustic waves by a circular disk [33], the diffraction of electromagnetic plane wave by a rectangular plate and a rectangular hole in conducting plate [34], and the diffraction of acoustic plane wave by a rectangular plate [35], [36]. The Kobayashi potential has also been used for the diffraction of electromagnetic waves by a thick slit [37], a flanged parallel-plate waveguide [38], an \emph{N}-slit array [39],etc.\vspace{12pt}
 \indent The Kobayashi potential method resembles the MoM in its spectrum domain, but the formulation is different. The MoM is based on an integral equation, whereas the Kobayashi potential method starts from the dual integral equation. The MoM in a space domain has been used mostly in the diffraction problems of electromagnetic waves. We can cite the following advantages of the Kobayashi potential method over the current numerical techniques (mainly over MoM).\vspace{12pt}
\begin{enumerate}
\item In contrast to the MoM in a space domain, the Kobayashi potential method does not involve singularities of the Green's functions, so we can obtain very accurate results. 
\item Since each function involved in the integrand of the potential functions satisfies a part of the required boundary condition, the convergence is very rapid. In this respect, the present method may be regarded as eigenfunction expansion of the geometries.
\item As in two-dimensional (2-D) problem, the Kobayashi potential methods may be applied to more complicated problem with related configurations. These problems may be formulated in a manner similar to the eigenfunction expansions in cylindrical and spherical geometries.
\item For 2-D problems, the solution to a two-slit diffraction can be used to predict the coupling between the slits asymptotically [40]. This is also expected in three-dimensional (3-D) problems.
\end{enumerate}
\indent The disadvantage is that, the tractable geometries of this method are limited to special shapes like rectangular and circular plates and their related geometries. A similar situation is seen for other conventional eigenfunction expansions.\\
\indent In this dissertation, in which we discuss the diffraction of electromagnetic waves by a circular disk and circular hole, the solution begins by introducing the Fourier sine and cosine transforms of the tangential components of the vector potentials. From the requirement of the boundary conditions on the plane exterior to the disk or hole, we obtained the dual integral equations for the transformed functions (or weighting function). The equation are solved by using the properties of the Weber-Schafheitlin discontinuous integrals. At this step, we can incorporate the required edge condition into the solution. The results include two kinds of arbitrary discrete parameters, so that the general solution is obtained by superposing these results. By imposing the remaining boundary conditions on the circular disk or on the circular hole, we have a matrix equation for the expansion coefficients. Matrix elements are given by an infinite integral, which are then expanded into infinite series, which is more convenient for numerical computation. The numerical results has been presented for the far-filed pattern and compared the results of the far-filed pattern with the corresponding physical optics (PO) method. More detailed and the mathematical formulation will be examined in the next chapter. 

\chapter{Diffraction of Electromagnetic Wave by Circular Disk and Circular Hole}

\paragraph{}The diffraction of electromagnetic plane wave by a circular disk and its complementary problem, diffraction by circular hole
in a perfectly conducting plane, are classical problems and has been formulated rigorously by using eigenfunction (spheroidal
function) expansion [24]$\sim$[26] and the method of the Kobayashi potential [28]. Meixner showed how to formulate the problem as the boundary value problem by using the rectangular
components of Hertz vector. He first split the Hertz vector into two parts. One is associated with the incident wave and other is related to the scattered
wave. He first noticed that the field on the circular aperture can be expressed in term of two dimensional scalar wave function
and expanded this function in term of the spheroidal function and he expressed it by taking into account of the assumed expression on the aperture. The auxiliary
function representing aperture field was determined by imposing the edge condition. By using this formula Meixner and Andrewski gave the numerical results.

\paragraph{}Following this idea Nomura and Katsura [28] reformulated this problem by using the Weber-Schafheitlin's discontinuous
integrals (Kobayashi Potential). We discuss here an alternative method to that by Nomura and Katsura. We have the longitudinal
components of the vector potentials of the electric and magnetic type and  drives the dual integral equations for the surface field
on the plane where disk or aperture is located. The solution for one of the pair equations, is expanded in terms of the functions which is derived by taking into account the Maxwell's equations, discontinuous
properties of the Weber-Schafheitlin's integral and the required edge conditions. The corresponding spectral functions of the
surface field may be derived by applying the vector Hankel transform discussed by Chew and Kong [41],[42]. This determines the weighting
functions of the vector potentials. The expansion coefficients are determined from the another of the dual equation by using the projection of the function space. Then
we can reduce the problem into matrix equation. Matrix elements are given by infinite integrals and can be expressed in term of infinite
series which is convenient for numerical computation. Since the formulation is more compact than the method by Nomura and
Katsura, it is promising to apply to more complex problem.

\section{Statement of the problem}

\paragraph{} The geometry of the problem and the associated coordinates are described in Fig.2.1, where radius of the hole and disk
is $a$ and thickness of the conducting plane is assumed to be negligibly small. There are two kind of polarizations for the incident
plane wave. Since we consider both scattering problem of disk and complementary circular aperture, we list the electromagnetic
filed components and the corresponding vector potentials $A_z$ and $F_z$ for the incident wave and wave reflected from the
infinite conducting plane located at $z=0$. The relation between $\textbf{(E,H)}$ and $(A_z,F_z)$ are given by

\begin{eqnarray}
 \textbf{H}&=&{1 \over \mu }\nabla\times A \\
\textbf{E}&=&-{1 \over \epsilon}\nabla\times F
\end{eqnarray}
 Since we are using the longitudinal components of the vector potentials of the electric and magnetics type, hence the above equations
can be rewritten as
\begin{eqnarray}
 \textbf{H}&=&{1\over\mu}\Bigl({\partial A_z \over \partial y} i_x-{\partial A_z \over \partial x}i_y\Bigr) = {1\over\mu}\Bigl({\partial A_z \over \rho \partial\phi}i_\rho-{\partial A_z \over \partial\rho}i_\phi\Bigr)\\
 \textbf{E}& =& -{1\over \epsilon}\Bigl({\partial F_z\over\partial y}i_x-{\partial F_z\over\partial x}i_y\Bigr) = -{1\over\epsilon}\Bigl({\partial F_z \over\rho\partial\phi}i_\rho-{\partial F_z\over\partial\rho}i_\phi \Bigr)
\end{eqnarray}
These relation can be used to find the corresponding electromagnetic field from the vector potentials.

\section{Plane Waves in Term of Cylindrical Wave Functions}

\paragraph{}To facilitate to enforce the boundary condition we express the plane wave in the circular coordinates by using the
generating function for Bessel function, given by
\begin{eqnarray}
\exp\Bigl[{x\over2}\Bigl(t-{1\over t}\Bigr)\Bigr]=\sum_{m=-\infty}^{\infty}J_m(x)t^m
\end{eqnarray}
Putting $t=\exp(j{\pi\over2}-j\phi)$ in above equation, we can write
\begin{eqnarray}
\exp[jx\cos\phi]=\sum_{m=-\infty}^\infty j^m J_m(x)\exp[-jm\phi]
\end{eqnarray}
Using the recurrence relation for the Bessel function, that is
$$  J_{-m}(x)=(-1)^m J_m(x) $$
Eq.(2.6) can be written as
\begin{eqnarray}
 \exp[j\kappa\rho\sin\theta_0\cos\phi]&=&\sum_{m=-\infty}^\infty j^m J_m(\kappa\rho\sin\theta_0)\exp(-jm\phi)\nonumber\\
& &= \sum_{m=0}^\infty \epsilon_m j^m J_m(\kappa\rho\sin\theta_0)cos m\phi
\end{eqnarray}
where $\epsilon_m$ is Neumann's constant defined by $\epsilon_m=1$ for $m=0$ and $\epsilon_m=2$ for $m\geq 1$, and the symbol $
\theta_0$ denotes the polar angle of the incident wave.

\section{Perpendicular (Horizontal or E) Polarization}

\paragraph{}Let us assume that the plane of incidence lies in $xz$ plane, then the incident electromagnetic plane wave on perfectly conducting plane
are given by
\begin{eqnarray}
\textbf{E}^i&=&i_yE_1\exp[j\kappa x\sin\theta_0+j\kappa z\cos\theta_0]\\
\textbf{H}^i&=&Y_0E_1(i_x\cos\theta_0-i_z\sin\theta_0)\exp[j\kappa x\sin\theta_0+j\kappa z\cos\theta_0]
\end{eqnarray}
Similarly the reflected fields can be expressed as
\begin{eqnarray}
\textbf{E}^r&=&-i_y E_1\exp[j\kappa x\sin\theta_0-j\kappa z\cos\theta_0]\\
\textbf{H}^r&=&Y_0E_1(i_x\cos\theta_0+i_z\sin\theta_0)\exp[j\kappa x\sin\theta_0-j\kappa z\cos\theta_0]
 \end{eqnarray}
 Using Eq.(2.3) and (2.4), the corresponding vector potential can written as
\begin{eqnarray}
  F_z^i&=&{\epsilon_0 E_1\over j\kappa \sin\theta_0}\exp[j\kappa x \sin\theta_0+j\kappa z \cos\theta_0]\\
  F_z^r&=&-{\epsilon_0 E_1\over j\kappa\sin\theta_0}\exp[j\kappa x \sin\theta_0-j\kappa z\cos\theta_0]
\end{eqnarray}
Using the transformation from rectangular to cylindrical coordinates, the tangential components of the electromagnetic field on the plane
$z=0$ are given by
\begin{eqnarray}
E_\rho&=&E_y\sin\phi=E_1\sin\phi\exp[j\kappa\rho\cos\phi\sin\theta_0]\\
H_\rho&=&H_x\cos\phi=Y_0E_1\cos\theta_0\cos\phi\exp[j\kappa \rho\cos\phi\sin\theta_0]\\
E_\phi&=&E_y\cos\phi=E_1\cos\phi\exp[j\kappa \rho\cos\phi\sin\theta_0]\\
H_\phi&=&-H_x\sin\phi=-Y_0E_1\cos\theta_0\cos\phi\exp[j\kappa\rho\cos\phi\sin\theta_0]
\end{eqnarray}
From the above equation, it can readily carried out
\begin{eqnarray}
H_\rho&=&Y_0E_\phi\cos\theta_0\\
H_\phi&=&-Y_0E_\rho\cos\theta_0
\end{eqnarray}
Now applying the wave transformation, Eq.(2.7), we have
\begin{eqnarray}
H_\rho^i&=&H_\rho^r=Y_0E_\phi^i\cos\theta_0\nonumber\\
& &=-jY_0E_1\cos\theta_0\sum_{m=0}^\infty\epsilon_m j^m J_m^\prime(\kappa\rho\sin\theta_0)
\cos m\phi\\
 H_\phi^i&=&H_\phi^r=-Y_0E_\rho^i\cos\theta_0\nonumber\\
& &=jY_0E_1\cos\theta\sum_{m=0}^\infty\epsilon_mj^m{m\over \kappa\rho\sin\theta_0}J_m(\kappa\rho\sin\theta_0)\sin m\phi
\end{eqnarray}
where we have used the relation,
\begin{eqnarray}
{x\over2}\Bigl(1+{1\over t}\Bigr)\exp\Bigl[{x\over 2}\Bigl(t-{1\over t}\Bigr)\Bigr]&=&\sum_{m=-\infty}^\infty m J_m(x)t^m\\
 {1\over 2}\Big(t-{1\over t}\Big)\exp\Big[{x\over 2}\Bigl(t-{1\over t}\Bigr)\Bigr]&=&\sum_{m=-\infty}^\infty J_m^\prime(x)t^m
\end{eqnarray}
where $J_m^\prime(x)$ is the derivative with respect to the argument and $Y_0$ is the intrinsic admittance of free space.

\section{Parallel (Vertical or H) Polarization}

\paragraph{}In this case, the corresponding incident and reflected waves are given by
\begin{eqnarray}
 \textbf{E}^i&=&E_2(i_x\cos\theta_0-i_z\sin\theta_0)\exp[j\kappa x \sin\theta_0+j\kappa z \cos\theta_0]\\
 \textbf{H}^i&=&-Y_0E_2i_y\exp[j\kappa x\sin\theta_0+j\kappa z\cos\theta_0]\\
 \textbf{E}^r&=&E_2(-i_x\cos\theta_0-i_z\sin\theta_0)\exp[j\kappa x\sin\theta_0-j\kappa z \cos\theta_0]\\
 \textbf{H}^r&=&-Y_0E_2i_y\exp[j\kappa x \sin\theta_0-j\kappa z\cos\theta_0]\\
 A_z^i&=&{\mu_0 Y_0 E_2\over j\kappa \sin\theta_0}\exp[j\kappa x\sin\theta_0+j\kappa z\cos\theta_0]\\
 A_z^r&=&{\mu_0 Y_0 E_2\over j \kappa\sin\theta_0}\exp[j\kappa x \sin\theta_0-j\kappa z\cos\theta_0]
\end{eqnarray}
and the cylindrical coordinate expression of these waves are
\begin{eqnarray}
 H_\rho^i&=&H_\rho^r=j Y_0E_2\sum_{m=0}^\infty \epsilon_m j^m{m\over \kappa \rho\sin\theta_0}J_m(\kappa\rho\sin\theta_0)\sin m\phi\\
 E_\phi^i&=&Z_0\cos\theta_0 H_\rho^i\\
 H_\phi^i&=&H_\phi^r=jY_0E_2\sum_{m=0}^\infty \epsilon_m j^m J_m^\prime(\kappa\rho\sin\theta_0)\cos m\phi\\
 E_\rho^i&=&-Z_0\cos\theta_0 H_\phi^i
\end{eqnarray}

\vfill
\bigskip
\begin{figure}[h]
\begin{center}
\includegraphics[width=15cm, height=8.5cm]{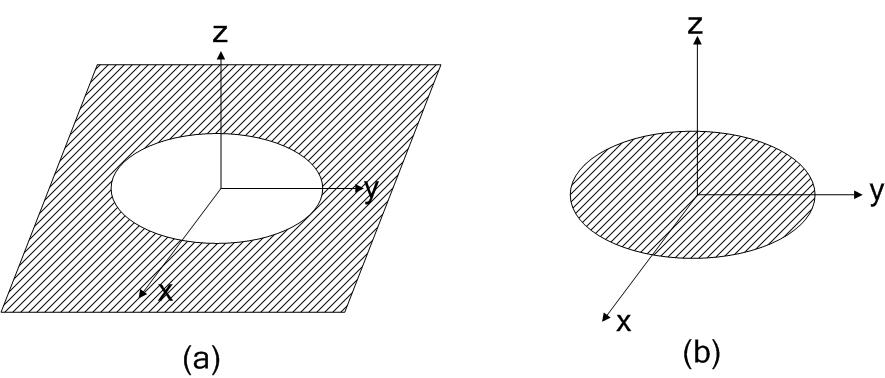}
\caption{\textbf{Diffraction of Plane Wave by a Circular Disk and a Circular Hole}\hfill}
\end{center}
\end{figure}
\vfill

\section{The Expression for the Fields Scattered by a Disk}
\paragraph{}We now discuss about our analytical method for predicting the filed scattered by a perfectly conducting disk on the plane at $z=0$.
We studied this problem by applying the Kobayashi potential $[45]$.

\section{Spectrum Function of the Current Density on the Disk}

\paragraph{} We assume the vector potential corresponding to the diffracted field as the superposition of the elementary solution for the wave function in the form
\begin{eqnarray}
 A_z^d(\rho,\phi,z)&=&\pm\mu_0 a\kappa Y_0\sum_{m=0}^\infty\int_0^\infty \Bigl[\widetilde f_{cm}(\xi)\cos m\phi+\widetilde f_{sm}(\xi)
\sin m\phi\Bigr]\nonumber\\
& &\times J_m(\rho_a\xi)\exp[\mp\sqrt{\xi^2-\kappa^2}z_a]\xi^{-1}d\xi\\
 F_z^d(\rho,\phi,z)&=&\epsilon_0 a\sum_{m=0}^\infty\int_0^\infty\Bigl[\widetilde g_{cm}(\xi)\cos m\phi+\widetilde g_{sm}(\xi)
\sin m\phi\Bigr]\nonumber \\
& &\times J_m(\rho_a\xi)\exp[\mp\sqrt{\xi^2-\kappa^2}z_a]\xi^{-1}d\xi
\end{eqnarray}
where the upper and lower signs refer to the region $z >0$ and $z <0$, respectively, and $\rho_a={\rho\over a}$ and $z_a={z\over a}$
are the normalized variables with respect to the radius $a$ of the disk. Also $ \widetilde f(\xi) $ and $ \widetilde g(\xi) $ are the unknown
spectrum functions and they are to be determined so that they satisfy all the required boundary condition. Eq.(2.34) and (2.35) are of the form of the Hankel transform
for $z=0$. It is seen that the tangential components of the electric filed derive from Eq.(2.34) and (2.35) satisfy the continuity on the plane $z=0$
due to the signs in front of Eq.(2.34). First we consider the surface field at the plane $z=0$ to derive the dual integral equations
associated with them. By using the relation between the vector potentials and the electromagnetic field, give by
\begin{eqnarray}
 \textbf{E}&=&-j\omega A-j{1\over\omega\mu\epsilon}\nabla(\nabla\cdot A)-{1\over\epsilon}\nabla\times F\\
 \textbf{H}&=&{1\over \mu}\nabla\times A-j\omega F-j{1\over\omega\mu\epsilon}\nabla(\nabla\cdot F)
\end{eqnarray}
Substituting Eq.(2.34) and (2.35) into the above equation, the electromagnetic field components tangential to $z-$ plane becomes
\begin{eqnarray}
E_\rho^d(\rho,\phi,0)&=&\sum_{m=0}^\infty\Bigl[E_{\rho c,m}(\rho_a)\cos m\phi+
E_{\rho s,m}(\rho_a)\sin m\phi\Bigr]\nonumber\\
& &=j\sum_{m=0}^\infty\int_0^\infty\sqrt{\xi^2-\kappa^2}\Bigl[\widetilde f_{cm}(\xi)\cos m\phi+\widetilde f_{sm}(\xi)\sin m\phi\Bigr]
J_m^\prime(\rho_a\xi)d\xi \nonumber\\
& &-\sum_{m=0}^\infty\int_0^\infty\Bigl[-\widetilde g_{cm}(xi)\sin m\phi+\widetilde g_{sm}(\xi)\cos m\phi\Bigr]{m\over\xi\rho_a}J_m(\rho_a\xi)d\xi \\
E_\phi^d(\rho,\phi,0)&=&\sum_{m=0}^\infty\Bigl[E_{\phi c,m}(\rho_a)\cos m\phi+E_{\phi s,m}(\rho_a)\sin m\phi\Bigr]\nonumber\\
& &=j\sum_{m=0}^\infty\int_0^\infty\sqrt{\xi^2-\kappa^2}\Bigl[-\widetilde f_{cm}(\xi)\sin m\phi+\widetilde f_{sm}\cos m\phi\Bigr]{m\over
\xi\rho_a}J_m(\rho_a\xi)d\xi\nonumber\\
& &+\sum_{m=0}^\infty\int_0^\infty\Big[\widetilde g_{cm}(\xi)\cos m\phi+\widetilde g_{sm}\sin m\phi\Bigr]J_m^\prime(\rho_a\xi)d\xi\\
H_\rho^d(\rho,\phi,0)&=&\sum_{m=0}^\infty\Bigl[H_{\rho c,m}(\rho_a)\cos m\phi+H_{\rho s,m}\sin m\phi\Bigr]\nonumber\\
& &=\pm\kappa Y_0\sum_{m=0}^\infty\int_0^\infty\Bigl[-\widetilde f_{cm}(\xi)\sin m\phi+\widetilde f_{sm}\cos m\phi\Bigr]{m\over\xi\rho_a}
J_m(\rho_a\xi)d\xi\nonumber\\
& &\pm{jY_0\over \kappa}\sum_{m=0}^\infty\int_0^\infty\sqrt{\xi^2-\kappa^2}\Bigl[\widetilde g_{cm}(\xi)\cos m\phi+\widetilde g_{sm}(\xi)\sin m\phi\Bigr]\nonumber\\
& &\times J_m^\prime(\rho_a\xi)d\xi\\
H_\phi^d(\rho,\phi,0)&=&\sum_{m=0}^\infty\Bigl[H_{\phi c,m}(\rho_a)\cos m\phi+H_{\phi s,m}(\rho_a)\sin m\phi\Bigr]\nonumber\\
& &=\mp \kappa Y_0\sum_{m=0}^\infty\int_0^\infty\Big[\widetilde f_{cm}(\xi)\cos m\phi+\widetilde f_{sm}\sin m\phi\Bigr]J_m^\prime(\rho_a\xi)d\xi\nonumber\\
& &\pm{jY_0\over \kappa}\sum_{m=0}^\infty\int_0^\infty\sqrt{\xi^2-\kappa^2}\Bigl[-\widetilde g_{cm}(\xi)\sin m\phi+\widetilde g_{sm}(\xi)
\cos m\phi\Bigr]\nonumber\\
& &\times{m\over \xi\rho_a}J_m(\rho_a\xi)d\xi
\end{eqnarray}
\indent In the above equations the upper and lower signs denote the values at $z=0_+$ and $z=0_-$, respectively. The tangential electric filed components take the same values at $z=0_+$ and $z=_-$. We rewrite Eq.(2.38)$\sim$(2.41) by using the matrices, given by
\begin{eqnarray}
\left[\matrix {E_{\rho c,m}(\rho_a) \cr E_{\phi s,m}(\rho_a)}\right]&=&\int_0^\infty\Bigl[H^-(\xi\rho_a)\Bigr]\left[\matrix{ j\sqrt{\xi^2-\kappa^2}\widetilde f_{cm}(\xi)\xi^{-1} \cr \widetilde g_{sm}(\xi)\xi^{-1}}\right]\xi d\xi\\
\left[\matrix {E_{\rho s,m}(\rho_a) \cr E_{\phi c,m}(\rho_a)}\right]&=&\int_0^\infty\Bigl[H^+(\xi\rho_a)\Bigr]\left[\matrix {j\sqrt{\xi^2-\kappa^2}\widetilde f_{sm}(\xi)\xi^{-1} \cr \widetilde g_{cm}(\xi)\xi^{-1}}\right]\xi d\xi
\end{eqnarray}
\begin{eqnarray}
\left[\matrix{ K_{\rho c,m}(\rho_a)\cr K_{\phi s,m}(\rho_a)}\right]&=&2Y_0\int_0^\infty\Bigl[H^-(\xi\rho_a)\Bigr]\left[\matrix {\kappa \widetilde f_{cm}(\xi)\xi^{-1} \cr j\sqrt{\xi^2-\kappa^2}\widetilde g_{sm}(\xi)(\kappa\xi)^{-1}}\right]\xi d\xi\nonumber\\
& &=\int_0^\infty\Bigl[H^-(\xi\rho_a)\Bigr]\left[\matrix {\widetilde K_{\rho c,m}(\xi)\cr \widetilde K_{\phi s,m}(\xi)}\right]\xi d\xi\\
\left[\matrix{K_{\rho s,m}(\rho_a)\cr K_{\phi c,m}(\rho_a)}\right]&=&2Y_0\int_0^\infty\Bigl[H^+(\xi\rho_a)\Bigr]\left[\matrix{\kappa \widetilde f_{sm}(\xi)\xi^{-1} \cr j\sqrt{\xi^2-\kappa^2}\widetilde g_{cm}(\xi)(\kappa\xi)^{-1}}\right]\xi d\xi\nonumber\\
& &=\int_0^\infty\Bigl[H^+(\xi\rho_a)\left[\matrix {\widetilde K_{\rho s,m}(\xi) \cr \widetilde K_{\phi c,m}(xi)}\right]\xi d\xi
\end{eqnarray}
where we use the current density $\textbf{K}$ instead of surface magnetic field by using the relation $K_\rho=-2H_\phi$ and $K_\phi=2H_\rho$. The kernel matrix $\Bigl[H^+(\xi\rho_a)\Bigr]$ and $\Bigl[H^-(\xi\rho_a)\Bigr]$ are defined by
\begin{eqnarray}
 \Bigl[H^\pm(\xi\rho_a)\Bigr]=\left[\matrix {J_m^\prime(\xi\rho_a)\qquad& \pm{m\over \xi\rho_a}J_m(\xi\rho_a)\cr\pm{m\over\xi\rho_a}J_m(\xi\rho_a)\qquad& J_m^\prime(\xi\rho_a)}\right]
\end{eqnarray}
\indent The required boundary conditions state that the current densities on the plane $z=0$ are zero for $\rho_a\geq 1$ and the tangential components of the total electric field vanish on the disk. These are written as
\begin{eqnarray}
\int_0^\infty\Bigl[H^-(\xi\rho_a)\Bigr]\left[\matrix{\widetilde K_{\rho c,m}(\xi)\cr \widetilde K_{\phi s,m}(\xi)}\right]\xi d\xi=0\qquad \rho_a\geq 1\\
\int_0^\infty\Bigl[H^+(\xi\rho_a)\Bigr]\left[\matrix{\widetilde K_{\rho s,m}(\xi)\cr \widetilde K_{\phi c,m}(\xi)}\right]\xi d\xi=0\qquad \rho_a\geq 1
\end{eqnarray}
\begin{eqnarray}
\left[\matrix{E^t_{\rho c,m}(\rho_a)\cr E^t_{\phi s,m}(\rho_a)}\right]&=&{Z_0\over 2}\int_0^\infty\Bigl[H^-(\xi\rho_a)\Bigr]\left[\matrix{ j\sqrt{\xi^2-\kappa^2}\widetilde f_{cm}(\xi)\xi^{-1}\cr\widetilde g_{sm}(\xi)\xi^{-1}}\right]\xi d\xi\nonumber\\
& &+\left[\matrix {E^i_{\rho c,m}(\rho_a)\cr E^i_{\phi s,m}(\rho_a)}\right]=0 \qquad\rho_a\leq 1 \\
\left[\matrix{E^t_{\rho s,m}(\rho_a)\cr E^t_{\phi c,m}(\rho_a)}\right]&=&{Z_0\over 2}\int_0^\infty\Bigl[H^+(\xi\rho_a)\Bigr]\left[\matrix{j\sqrt{\xi^2-\kappa^2}\widetilde f_{sm}(\xi)\xi^{-1}\cr \widetilde g_{cm}(\xi)\xi^{-1}}\right]\xi d\xi\nonumber\\
& &+\left[\matrix{E^i_{\rho s,m}(\rho_a)\cr E^i_{\phi c,m}(\rho_a)}\right]=0\qquad \rho_a\leq 1
\end{eqnarray}
where $E^i_{\rho c,m}$ and $E^i_{\rho s,m}$ denote the $\cos m\phi$ and $\sin m\phi$ parts of the incident wave $E^i_\rho$, respectively, and the same is true for $E^i_{\phi c,m}$ and $E^i_{\phi s,m}$. The expression for theses factor can be obtained, and is given by
\begin{eqnarray}
 E^i_{\rho c}(\rho_a)&=&-jE_2\cos\theta_0\epsilon_m j^m J_m^\prime(\kappa\rho_a\sin\theta_0)\\
 E^i_{\rho s}(\rho_a)&=&-jE_1\epsilon_m j^m {m\over \kappa\rho_a\sin\theta_0}J_m(\kappa\rho_a\sin\theta_0)\\
 E^i_{\phi c}(\rho_a)&=&-jE_1\epsilon_m j^m J_m^\prime(\kappa\rho_a\sin\theta_0)\\
 E^i_{\phi s}(\rho_a)&=&jE_2\cos\theta_0\epsilon_m j^m{m\over \kappa\rho_a\sin\theta_0}J_m(\kappa\rho_a\sin\theta_0)
\end{eqnarray}
Eq.(2.47)$\sim$(2.50) are the dual integral equations to determine the spectrum functions. To solve these equations, we expand $\textbf{K}(\rho_a)$ by the function which satisfy the Maxwell's equations and the edge conditions. These functions can be found by taking into account the discontinuity property of the Weber-Schafheitlin's integrals. Once the expression for $\textbf{K}(\rho_a)$ are established, the corresponding spectrum functions can be derived by applying the vector Hankel transform introduced by Chew and Kong [41], [42]. Using the vector Hankel transform, [see Appendix A], Eq.(2.44) and (2.45) can be rewritten as
\begin{eqnarray}
\left[\matrix {\widetilde K_{\rho_ c,m}(\xi) \cr \widetilde K_{\phi s,m}(\xi)}\right]=\int_0^\infty \Bigl[H^-(\xi\rho_a)\Bigr]\left[\matrix { K_{\rho c,m}(\rho_a)\cr K_{\phi s,m}(\rho_a)}\right]\rho_a d\rho_a\\
 \left[\matrix {\widetilde K_{\rho s,m}(\xi)\cr \widetilde K_{\phi c,m}(\xi)}\right]=\int_0^\infty\Bigl[H^+(\xi\rho_a)\Bigr]\left[\matrix{ K_{\rho s,m}(\rho_a)\cr K_{\phi c,m}(\rho_a)}\right]\rho_a d\rho_a
\end{eqnarray}
\indent It is noted that $(K_\rho,K_\phi)$ satisfy the vector Helmholtz equation $\nabla^2 \textbf{K}+\kappa^2 \textbf{K}=0$ in circular cylindrical coordinates, since $\textbf{K}$ and $\textbf{H}$ are relate by $\textbf{K}=n\times\textbf{ H}$ on the plane $z=0$. Furthermore $(K_\rho,K_\phi)$ have the property $K_\rho\sim(1-\rho_a^2)^{{1\over2}}$ and $K_\phi\sim(1-\rho_a^2)^{-{1\over2}}$ near the edge of the disk. By taking into these facts, we substituting the Kernel matrix in Eq.(2.55), we can write

\begin{eqnarray}
\int_0^\infty \left[\matrix {J_m^\prime(\xi\rho_a) \quad & -{m\over \xi\rho_a}J_m(\xi\rho_a)\cr -{m\over\xi\rho_a}J_m(\xi\rho_a)\quad & J_m^\prime(\xi\rho_a)}\right]\left[\matrix {\widetilde K_{\rho c,m}(\xi)\cr \widetilde K_{\phi s,m}(\xi)}\right]\xi d\xi=0
\end{eqnarray}
Using the recurrence relation for Bessel function, given by
$$ J_{m-1}(x)+J_{m+1}(x)={m\over x}J_m(x)$$
and
$$ J_{m-1}(x)-J_{m+1}(x)=J_m^\prime(x)$$
and the properties of Weber Schafheitlin's discontinuous integral, [see Appendix B] , Eq.(2.47)and (2.48) can be satisfied if we choose,
\begin{eqnarray}
K_{\rho c,m}(\rho_a)&=&\sum_{n=0}^\infty\Bigl[A_{mn}F^-_{mn}(\rho_a)-B_{mn}G^+_{mn}(\rho_a)\Bigr]\\
K_{\rho s,m}(\rho_a)&=&\sum_{n=0}^\infty\Bigl[C_{mn}F^-_{mn}(\rho_a)+B_{mn}G^+_{mn}(\rho_a)\Bigr]\\
K_{\phi s,m}(\rho_a)&=&\sum_{n=0}^\infty\Bigl[-A_{mn}F^+_{mn}(\rho_a)+B_{mn}G^-_{mn}(\rho_a)\Bigr]\\
K_{\phi c,m}(\rho_a)&=&\sum_{n=0}^\infty\Bigl[C_{mn}F^+_{mn}(\rho_a)+D_{mn}G^-_{mn}(\rho_a)\Bigr]
\end{eqnarray}
where
\begin{eqnarray}
F^\pm_{mn}(\rho_a)=\int_0^\infty\Bigl[J_{|m-1|}(\eta\rho_a)J_{|m-1|+2n+{1\over 2}}(\eta)\pm J_{m+1}(\eta\rho_a)J_{m+2n+{3\over 2}}(\eta)\Bigr]\eta^{{1\over 2}}d\eta\\
G^\pm_{mn}(\rho_a)=\int_0^\infty\Bigl[J_{|m-1|}(\eta\rho_a)J_{|m-1|+2n+{3\over 2}}(\eta)\pm J_{m+1}(\eta\rho_a)J_{m+2n+{5 \over 2}}(\eta)\Bigr]\eta^{-{1\over 2}}d\eta
\end{eqnarray}
\indent It may readily be verified that $F^\pm_{mn}(\rho_a)=G^\pm_{mn}(\rho_a)=0$ for $\rho_a\geq1$, and $F^+_{mn}(\rho_a)\sim(1-\rho_a^2)^{-{1\over2}}$, $F^-_{mn}(\rho_a)\sim(1-\rho_a^2)^{{1\over2}}$, $G^+_{mn}(\rho_a)\sim(1-\rho_a^2)^{{1\over2}}$ and $G^-_{mn}(\rho_a)\sim(1-\rho_a^2)^{{3\over2}}$ near the edge $\rho_a\simeq1$. Thus the above expressions satisfy one part of the dual integral equations with the unknown expansion coefficients $A_{mn}\sim D_{mn}$. To derive the spectrum functions $\widetilde f(\xi)$ and $\widetilde g(\xi)$ of the vector potentials, we first determine the spectrum functions of the current densities, since they are related to each other. Now substituting Eq.(2.58)$\sim$(2.61) into Eq.(2.55) and (2.56) and perform the integration, then the spectrum function of the current density is determined. The results are
\begin{eqnarray}
\widetilde K_{\rho c,m}(\xi)&=&\sum_{n=0}^\infty\Bigl[A_{mn}\Xi^+_{mn}(\xi)-B_{mn}\Gamma^-_{mn}(\xi)\Bigr]\\
\widetilde K_{\phi s,m}(\xi)&=&\sum_{n=0}^\infty\Bigl[-A_{mn}\Xi^-_{mn}(\xi)+B_{mn}\Gamma^+_{mn}(\xi)\Bigr]\\
\widetilde K_{\rho s,m}(\xi)&=&\sum_{n=0}^\infty\Bigl[C_{mn}\Xi^+_{mn}(\xi)+D_{mn}\Gamma^-_{mn}(\xi)\Bigr]\\
\widetilde K_{\phi c,m}(\xi)&=&\sum_{n=0}^\infty\Bigl[C_{mn}\Xi^-_{mn}(\xi)+D_{mn}\Gamma^+_{mn}(\xi)\Bigr]
\end{eqnarray}
for $m\geq 1$ and
\begin{eqnarray}
\widetilde K_{\rho c,0}(\xi)&=&2\sum_{n=0}^\infty B_{0n}J_{2n+{5\over 2}}(\xi)\xi^{-{3\over 2}}\\
\widetilde K_{\phi s,0}(\xi)&=&2\sum_{n=0}^\infty A_{0m}J_{2n+{3\over 2}}(\xi)\xi^{-{1\over 2}}\\
\widetilde K_{\rho s,0}(\xi)&=&-2\sum_{n=0}^\infty D_{0n}J_{2n+{5\over 2}}(\xi)\xi^{-{3\over 2}}\\
\widetilde K_{\phi c,0}(\xi)&=&-2\sum_{n=0}^\infty C_{0n}J_{2n+{3\over 2}}(\xi)\xi^{-{1\over 2}}
\end{eqnarray}
for $m=0$.
In the above equations the functions $\Xi^\pm_{mn}(\xi)$ and $\Gamma^\pm_{mn}(\xi)$ are defined by
\begin{eqnarray}
\Xi_{mn}^\pm(\xi)=\Bigl[J_{m+2n-{1\over 2}}(\xi)\pm J_{m+2n+{3\over 2}}\Bigr]\xi^{-{1\over 2}}\\
\Gamma_{mn}^\pm(\xi)=\Bigl[J_{m+2n+{1\over 2}}(\xi)\pm J_{m+2n+{5\over 2}}(\xi)\Bigr]\xi^{-{3\over 2}}
\end{eqnarray}
In deriving Eq.(2.64)$\sim$(2.67) and Eq.(2.68)$\sim$(2.71), we used the formula of the Hankel transform given by
$$\int_0^\infty J_m(\alpha\rho)J_m(\beta\rho)\rho d\rho={\delta(\alpha-\beta)\over \alpha}$$
and the property of the delta function given by
$$\int_0^\infty f(x)\delta(x-x^\prime)dx=f(x^\prime)$$
\indent From Eq.(2.44) and (2.45), the spectral functions $\widetilde f_{cm}(\xi)\sim\widetilde g_{sm}(\xi)$ can be expressed in term of $\widetilde K_{\rho cs}(\xi)\sim\widetilde K_{\phi cs}(\xi)$.
\begin{eqnarray}
\widetilde f_{cm}(\xi)&=&{Z_0\over 2\kappa}\widetilde K_{\rho c,m}(\xi)\xi\\
\widetilde f_{sm}(\xi)&=&{Z_0\over 2\kappa}\widetilde K_{\rho s,m}(\xi)\xi\\
\widetilde g_{cm}(\xi)&=&{\kappa Z_0\over j2\sqrt{\xi^2-\kappa^2}}\widetilde K_{\phi c,m}(\xi)\xi\\
\widetilde g_{sm}(\xi)&=&{\kappa Z_0\over j2\sqrt{\xi^2-\kappa^2}}\widetilde K_{\phi s,m}(\xi)\xi
\end{eqnarray}
These expression gives the relation between the weight function and the spectrum function of the current density, which can be used for the derivation of the expansion coefficient.

\section{Derivation of the Expansion Coefficients}

\paragraph{}The equation for the expansion coefficients can be obtained by applying the remaining boundary condition that the tangential components of the electric field vanish on the disk, which is given by Eq.(2.49) and (2.50). By substituting Eq.(2.74)$\sim$(2.77) into Eq.(2.49) and (2.50) , we have the relation given by
\begin{eqnarray}
\left[\matrix{E^t_{\rho c,m}(\rho_a)\cr E^t_{\phi s,m}(\rho_a)}\right]&=&{Z_0\over 2}\int_0^\infty\Bigl[H^-(\xi\rho_a)\Bigr]\left[\matrix{{j\sqrt{\xi^2-\kappa^2}\over\kappa}\widetilde K_{\rho c,m}\cr{\kappa\over j\sqrt{\xi^2-\kappa^2}}\widetilde K_{\phi s,m}(\xi)}\right]\xi d\xi\nonumber\\
& &+\left[\matrix{E^i_{\rho c,m}(\rho_a)\cr E^i_{\phi s,m}(\rho_a)}\right]=0 \qquad \rho_a\leq 1\\
\left[\matrix{E^t_{\rho s,m}(\rho_a)\cr E^t_{\phi c,m}(\rho_a)}\right]&=&{Z_0\over 2}\int_0^\infty\Bigl[H^+(\xi\rho_a)\left[\matrix{{j\sqrt{\xi^2-\kappa^2}\over \kappa}\widetilde K_{\rho s,m}(\xi)\cr {\kappa\over j\sqrt{\xi^2-\kappa^2}}\widetilde K_{\phi c,m}(\xi)}\right]\xi d\xi\nonumber\\
& & +\left[\matrix{E^i_{\rho s,m}(\rho_a)\cr E^i_{\phi c,m}(\rho_a)}\right]=0\qquad \rho_a\leq 1
\end{eqnarray}
\indent Eq.(2.78) and (2.79 ) are projected into function space with element $ v_n^m(\rho_a^2)$ for $ E_\rho $ and $u_n^m(\rho_a^2)$ for $E_\phi$, [see Appendix c], then we obtain the matrix equations for the expansion coefficients $A_{mn}\sim D_{mn}$. The results are given by
@finalout
\begin{eqnarray}
@finalout
\sum_{n=0}^\infty A_{mn}\left \{{j\over\kappa}\int_0^\infty{\sqrt{\xi^2-\kappa^2}\over \xi}\Bigl[J_{m+2n-{1\over 2}}(\xi)+J_{m+2n+{3\over 2}}(\xi)\Bigr] \Bigl[\alpha_p^m J_{m+2p+{1\over 2}}(\xi)-(\alpha_p^m+3)J_{m+2p+{5\over 2}}(\xi)\Bigr]d\xi\right.\nonumber\vspace{12pt}
\left. -j\kappa m\int_0^\infty{1\over \xi\sqrt{\xi^-\kappa^2}}\Bigl[J_{m+2n-{1\over 2}}(\xi)-J_{m+2n+{3\over 2}}(\xi)\Bigr]\Bigl[J_{m+2p+{1\over 2}}(\xi)+J_{m+2p+{5\over 2}}(\xi)\Bigr]d\xi \right\}\nonumber\vspace{12pt}
-\sum_{n=0}^\infty B_{mn}\left\{{j\over \kappa}\int_0^\infty{\sqrt{\xi^2-\kappa^2}\over \xi^2}\Bigl[J_{m+2n+{1\over 2}}(\xi)-J_{m+2n+{5\over 2}}(\xi)\Bigr]\Bigl[\alpha^m_p J_{m+2p+{1\over 2}}(\xi)-(\alpha^m_p+3) J_{m+2p+{5\over 2}}(\xi)\Bigr]d\xi\right.\nonumber\vspace{12pt}
\left.+j\kappa m\int_0^\infty{1\over \xi^2\sqrt{\xi^2-\kappa^2}}\Bigl[J_{m+2n+{1\over 2}}(\xi)+J_{m+2n+{5\over 2}}(\xi)\Bigr]\Bigl[J_{m+2p+{1\over 2}}(\xi)+J_{m+2p+{5\over 2}}(\xi)\Bigr]d\xi\right\}\nonumber\vspace{12pt}
=j4Y_0E_2\cos\theta_0 j^m\Bigl[\alpha^m_p J_{m+2p+{1\over 2}}(\kappa\sin\theta_0)-(\alpha^m_p+3)J_{m+2p+{5\over 2}}(\kappa\sin\theta_0)\Bigr](\kappa\sin\theta_0)^{-{3\over2}}\nonumber\vspace{12pt}
 m=1,2,3,\cdots ,\quad  p=0,1,2,3,\cdots ,\nonumber\vspace{12pt}
\end{eqnarray}
\begin{eqnarray}
\sum_{n=0}^\infty A_{mn}\left\{-{jm\over \kappa}\int_0^\infty\sqrt{\xi^2-\kappa^2}\Bigl[J_{m+2n-{1\over 2}}(\xi)+J_{m+2n+{3\over 2}}(\xi)\Bigr]\Bigl[J_{m+2p-{1\over 2}}(\xi)-J_{m+2p+{3\over 2}}(\xi)\Bigr]d\xi\right.\nonumber\\
\left.+j\kappa\int_0^\infty {1\over \sqrt{\xi^2-\kappa^2}}\Bigl[J_{m+2n-{1\over 2}}(\xi)-J_{m+2n+{3\over 2}}(\xi)\Bigr]\Bigl[\alpha^m_p J_{m+2p-{1\over 2}}(\xi)-(\alpha^m_p+1)J_{m+2p+{3\over 2}}(\xi)\Bigr]d\xi\right\}\nonumber\\
+\sum_{n=0}^\infty B_{mn}\left\{{jm\over\kappa}\int_0^\infty{\sqrt{\xi^2-\kappa^2}\over \xi}\Bigl[J_{m+2n+{1\over 2}}(\xi)-J_{m+2n+{5\over 2}}(\xi)\Bigr]\Bigl[J_{m+2p-{1\over 2}}(\xi)+J_{m+2p+{3\over 2}}(\xi)\Bigr]d\xi\right.\nonumber\\
\left. -j\kappa\int_0^\infty {1\over \xi \sqrt{\xi^2-\kappa^2}}\Bigl[J_{m+2n+{1\over 2}}(\xi)+J_{m+2n+{5\over 2}}(\xi)\Bigr]\Bigl[\alpha^m_p J_{m+2p-{1\over 2}}(\xi)-(\alpha^m_p+1)J_{m+2p+{3\over 2}}(\xi)\Bigr]d\xi\right\}\nonumber\\
=-j4Y_0E_2\cos\theta_0 j^m m\Bigl[J_{m+2p-{1\over 2}}(\kappa\sin\theta_0)+J_{m+2p+{3\over 2}}(\kappa\sin\theta_0)\Bigr](\kappa\sin\theta_0)^{-{1\over 2}}\nonumber\\
\qquad m=1,2,3,\cdots , \quad p=0,1,2,3,\cdots ,\nonumber \\
\end{eqnarray}
\begin{eqnarray}
\sum_{n=0}^\infty C_{mn}\left\{{j\over \kappa}\int_0^\infty{\sqrt{\xi^2-\kappa^2}\over \xi}\Bigl[J_{m+2n-{1\over 2}}(\xi)+J_{m+2n+{3\over 2}}(\xi)\Bigr]\Bigl[\alpha^p_m J_{m+2p+{1\over 2}}(\xi)-(\alpha^m_p+3)J_{m+2p+{5\over 2}}(\xi)\Bigl]d\xi\right.\nonumber\\
\left. -j\kappa m\int_0^\infty{1\over \xi\sqrt{\xi^2-\kappa^2}}\Bigl[J_{m+2n-{1\over 2}}(\xi)-J_{m+2n+{3\over 2}}(\xi)\Bigr]\Bigl[J_{m+2p+{1\over 2}}(\xi)+J_{m+2p+{5\over 2}}(\xi)\Bigr]d\xi\right\}\nonumber\\
+\sum_{n=0}^\infty D_{mn}\left\{{j\over \kappa}\int_0^\infty{\sqrt{\xi^2-\kappa^2}\over \xi^2}\Bigl[J_{m+2n+{1\over 2}}(\xi)-J_{m+2n+{5\over 2}}(\xi)\Bigr]\Bigl[\alpha^m_pJ_{m+2p+{1\over 2}}(\xi)-(\alpha^m_p+3)J_{m+2p+{5\over 2}}(\xi)\Bigr]d\xi\right.\nonumber\\
\left. -j\kappa m\int_0^\infty {1\over \xi^2\sqrt{\xi^2-\kappa^2}}\Bigl[J_{m+2n+{1\over 2}}(\xi)+J_{m+2n+{5\over 2}}(\xi)\Bigr]\Bigl[ J_{m+2p+{1\over 2}}(\xi)+J_{m+2p+{5\over 2}}(\xi)\Bigr]d\xi\right\}\nonumber\\
=j4Y_0E_1 j^m m\Bigl[J_{m+2p+{1\over 2}}(\kappa\sin\theta_0)+J_{m+2p+{5\over 2}}(\kappa\sin\theta_0)\Bigr](\kappa\sin\theta_0)^{-{3\over 2}}\nonumber\\
\qquad m=1,2,3,\cdots , \quad p=0,1,2,3,\cdots ,\nonumber\\
\end{eqnarray}
\begin{eqnarray}
 \sum_{n=0}^\infty C_{mn}\left\{{jm\over \kappa}\int_0^\infty \sqrt{\xi^2-\kappa^2}\Bigl[J_{m+2n-{1\over 2}}(\xi)+J_{m+2n+{3\over 2}}(\xi)\Bigr]\Bigl[J_{m+2p-{1\over 2}}(\xi)+J_{m+2p+{3\over 2}}(\xi)\Bigr]d\xi\right.\nonumber\\
\left. -j\kappa\int_0^\infty{1\over \sqrt{\xi^2-\kappa^2}}\Bigl[J_{m+2n-{1\over 2}}(\xi)-J_{m+2n+{3\over 2}}(\xi)\Bigr]\Bigl[\alpha^m_p J_{m+2p-{1\over 2}}(\xi)-(\alpha^m_p+1)J_{m+2p+{3\over 2}}(\xi)\Bigr]d\xi\right\}\nonumber\\
+\sum_{n=0}^\infty D_{mn}\left\{{jm\over \kappa}\int_0^\infty {\sqrt{\xi^2-\kappa^2}\over \xi}\Bigl[J_{m+2n+{1\over 2}}(\xi)-J_{m+2n+{5\over 2}}(\xi)\Bigr]\Bigl[J_{m+2p-{1\over 2}}(\xi)+J_{m+2p+{3\over 2}}(\xi)\Bigr]d\xi\right.\nonumber\\
\left. -j\kappa\int_0^\infty {1\over \xi\sqrt{\xi^2-\kappa^2}}\Bigl[J_{m+2n+{1\over 2}}(\xi)+J_{m+2n+{5\over 2}}(\xi)\Bigr]\Bigl[\alpha^m_p J_{m+2p-{1\over 2}}(\xi)-(\alpha^m_p+1)J_{m+2p+{3\over 2}}(\xi)\Bigr]d\xi\right\}\nonumber\\
=j4Y_0E_1j^m\Bigl[\alpha^m_pJ_{m+2p-{1\over 2}}(\kappa\sin\theta_0)-(\alpha^m_p+1)J_{m+2p+{3\over 2}}(\kappa\sin\theta_0)\Bigr](\kappa\sin\theta_0)^{-{1\over 2}}\nonumber\\
\qquad m=1,2,3,\cdots ,\quad p=0,1,2,3,\cdots ,\nonumber\\
\end{eqnarray}

\begin{eqnarray*}
\sum_{n=0}^\infty B_{0n}{1\over\kappa}\int_0^\infty{\sqrt{\xi^2-\kappa^2}\over \xi^2} J_{2n+{5\over 2}}(\xi)\Bigl[-pJ_{2p+{1\over 2}}(\xi)+(p+1.5)J_{2p+{5\over 2}}(\xi)\Bigr]d\xi\nonumber\\
=Y_0E_2\cos\theta_0\Bigl[-pJ_{2p+{1\over 2}}(\kappa\sin\theta_0)+(p+1.5)J_{2p+{5\over 2}}(\kappa\sin\theta_0)\Bigr](\kappa\sin\theta_0)^{-{3\over 2}}\nonumber\\
\end{eqnarray*}

\begin{eqnarray}
\sum_{n=0}^\infty C_{0n}\kappa\int_0^\infty{1\over \sqrt{\xi^2-\kappa^2}}J_{2n+{3\over 2}}(\xi)\Bigl[-pJ_{2p-{1\over 2}}(\xi)+(p+1.5)J_{2p+{3\over 2}}(\xi)\Bigr]d\xi\nonumber\\
=Y_0E_1\Bigl[-pJ_{2p-{1\over 2}}(\kappa\sin\theta_0)+(p+0.5)J_{2p+{3\over 2}}(\kappa\sin\theta_0)\Bigr](\kappa\sin\theta_0)^{-{1\over 2}}\nonumber\\
\qquad p=01,2,3,\cdots ,\nonumber\\
\end{eqnarray}

where $\alpha^m_p=m+2p$. From Eq.(2.80)$\sim$(2.84) we find that all of the matrix elements have the form $K(\alpha,\beta\lambda)$ or $G(\alpha,\beta,\lambda)$, which are defined by
\begin{eqnarray}
 G(\alpha,\beta;\kappa)&=&\int_o^\infty{J_\alpha(\xi)J_\beta(\xi)\over\sqrt{\xi^2-\kappa^2}}d\xi\nonumber\\
 G_2(\alpha,\beta,\lambda)&=&\int_0^\infty {1\over \xi^\lambda\sqrt{\xi^2-\kappa^2}}J_\alpha(\xi) J_\beta(\xi)d\xi\\
 K(\alpha,\beta,\lambda)&=&\int_0^\infty{\sqrt{\xi^2-\kappa^2}\over \xi^\lambda}J_\alpha(\xi)J_\beta(\xi)d\xi
\end{eqnarray}

\indent These integral converge when $\alpha+\beta > -1$ and $\lambda > 1$ for $K(\alpha,\beta,\lambda)$ and $ \alpha+\beta >  \lambda-1 $ and $\lambda > -1$ for $ G(\alpha,\beta,\lambda)$. By using the recurrence relation for the Bessel function, we can find that all the matrix elements contained in Eq.(2.80)$\sim$(2.84) converge for all indices. It is worthwhile to note that $K(\alpha,\beta,\lambda)$ and $G(\alpha,\beta,\lambda)$ can be integrated into infinite series which are more convenient for  numerical computation.

\section{Far Field Expression}

\paragraph{}The expression for the far filed can be drive by two different way. One method is to evaluate the field radiated from the current density induced on the disk and the second method is to evaluate the expression of the vector potentials given in Eq.(2.34) and (2.35), directly by applying the stationary phase method of integration. Let the rectangular components of the current density be $K_x(\rho_a^\prime,\phi^\prime)$ and $K_y(\rho_a^\prime,\phi^\prime)$, then the vector potential produce by this current density is given by
\begin{eqnarray}
 A_x&=&{\mu_0\over 4\pi}\int_S K_x(\rho_a^\prime,\phi^\prime){\exp(-j\kappa r)\over r}ds^\prime\\
 A_y&=&{\mu_0\over 4\pi}\int_S K_y(\rho_a^\prime,\phi^\prime){\exp(-j\kappa r)\over r}ds^\prime
\end{eqnarray}
where $r$ is the radial distance between the observation and source points.\\
 \indent Since we are interested in the far field, it can be shown that the radial distance $r$  from any point on the source or scatterer to the observation point can be assumed to be parallel to the radial distance $R$ from the origin to the observation point. In such cases the relation between the magnitude of $r$ and $R$ , given by
$$ r=\sqrt{R^2-2R\rho^\prime\sin\theta\cos(\phi-\phi^\prime+\rho^{\prime 2})}$$
can be approximated, most commonly by [43]
\[r\simeq\left\{
\begin{array}{l l}
R-\rho^\prime\sin\theta\cos(\phi-\phi^\prime)&\quad \mbox{for phase variations}\\R&\quad\mbox{for amplitude variation}\\
\end{array}\right.\]
Substituting the above equation into Eq.(2.87)and (2.88), we have
\begin{eqnarray}
A_x&=&\mu_0 G_0(R)\int_S K_x(\rho_a^\prime,\phi^\prime)\exp[j\kappa\rho_a^\prime\sin\theta\cos(\phi-\phi^\prime)]dS^\prime\\
A_y&=&\mu_0G_0(R)\int_S K_y(\rho_a^\prime,\phi^\prime)\exp[j\kappa\rho_a^\prime\sin\theta\cos(\phi-\phi^\prime)]dS^\prime
\end{eqnarray}
where $ G_0(R)$ is defined by
$$G_0(R)={\exp(-j\kappa R)\over 4\pi R}$$
Using the transformation from cylindrical-to-rectangular coordinates, the rectangular components of the current density can be express as
\begin{eqnarray}
K_x(\rho_a^\prime,\phi^\prime)&=&K_\rho\cos\phi^\prime-K_\phi\sin\phi^\prime\\
K_y(\rho_a^\prime,\phi^\prime)&=&K_\rho\sin\phi^\prime+K_\phi\cos\phi^\prime
\end{eqnarray}
Since,
\begin{eqnarray}
 K_\rho&=&\sum_{m=0}^\infty\Bigl[K_{\rho c,m}\cos m\phi^\prime+K_{\rho s,m}\sin m\phi^\prime\Bigr]\\
 K_\phi&=&\sum_{m=0}^\infty\Bigl[K_{\phi c,m}\cos m\phi^\prime+K_{\phi s,m}\sin m \phi^\prime\Bigr]
\end{eqnarray}
Hence,
\begin{eqnarray}
K_x(\rho_a^\prime,\phi^\prime)&=&\sum_{m=0}^\infty\Bigl[\Bigl(K_{\rho c,m}\cos m\phi^\prime+K_{\rho s,m}\sin m\phi\Big)\cos\phi^\prime-\Bigl(K_{\phi c,m}\cos m\phi^\prime\nonumber\\
& &+K_{\phi s,m}\sin m\phi^\prime\Bigr)\sin\phi^\prime\Bigr]\\
K_y(\rho_a^\prime,\phi^\prime)&=&\sum_{m=0}^\infty\Bigl[\Bigl(K_{\rho c,m}\cos m\phi^\prime+K_{\rho s,m}\sin m\phi^\prime\Bigr)\sin\phi^\prime+\Bigl(K_{\phi c,m}\cos m\phi^\prime\nonumber\\
& &+K_{\phi s,m}\sin m\phi^\prime\Bigr)\cos\phi^\prime\Bigr]
\end{eqnarray}
Substituting Eq.(2.91) and (2.92) into the above equation, we can write
\begin{eqnarray}
K_x(\rho^\prime,\phi^\prime)&=&{1\over 2}\sum_{m=0}^\infty\left\{\Bigl(K_{\rho c,m}-K_{\phi s,m}\Bigr)\cos[(m-1)\phi^\prime]+\Bigl(K_{\rho c,m}+K_{\phi s,m}\Bigr)\right.\nonumber\\
& &\left.\times\cos[(m+1)\phi^\prime]\right.\nonumber\\
& &\left.+\Bigl(K_{\rho s,m}+K_{\phi c,m}\Bigr)\sin[(m-1)\phi^\prime]+\Bigl(K_{\rho s,m}-K_{\phi c,m}\Bigr)\right.\nonumber\\
& &\left.\times\sin[(m+1)\phi^\prime]\right\}\\
K_y(\rho^\prime,\phi^\prime)&=&{1\over 2}\sum_{m=0}^\infty\left\{-\Bigl(K_{\rho c,m}-K_{\phi s,m}\Bigr)\sin[(m-1)\phi^\prime]+\Bigl(K_{\rho c,m}+K_{\phi s,m}\Bigr)\right.\nonumber\\
& &\left.\times\sin[(m+1)\phi^\prime]\right.\nonumber\\
& &\left.+\Bigl(K_{\rho s,m}+K_{\phi c,m}\Bigr)\cos[(m-1)\phi^\prime]-\Bigl(K_{\rho s,m}-K_{\phi c,m}\Bigr)\right.\nonumber\\
& &\left.\times\cos[(m+1)\phi^\prime]\right\}
\end{eqnarray}
where we have use the relation,
$$\cos\alpha\cos\beta={1\over 2}\cos(\alpha-\beta)+{1\over 2}\cos(\alpha+\beta)$$
$$\sin\alpha\sin\beta={1\over 2}\cos(\alpha-\beta)-{1\over 2}\cos(\alpha+\beta)$$
$$\sin\alpha\cos\beta={1\over 2}\sin(\alpha-\beta)+{1\over 2}\sin(\alpha+\beta)$$
Now substituting Eq.(2.97) and (2.98) into Eq.(2.89) and (2.90), we can write
\begin{eqnarray}
A_x&=&\pi a^2\mu_0G_0(R)\sum_{m=0}^\infty j^{m-1}\int_0^a \rho_a^\prime d\rho_a^\prime\nonumber\\
& &\times\left\{\Bigl(K_{\rho c,m}-K_{\phi s,m}\Bigr)J_{m-1}(\kappa\rho_a^\prime\sin\theta)\cos[(m-1)\phi]-\Bigl(K_{\rho c,m}+K_{\phi s,m}\Bigr)\right.\nonumber\\
& &\left.\times J_{m+1}(\kappa\rho_a^\prime\sin\theta)\cos[(m+1)\phi]\right.\nonumber\\
& &\left. +\Bigl(K_{\rho s,m}+K_{\phi c,m}\Bigr)J_{m-1}(\kappa\rho_a^\prime\sin\theta)\sin[(m-1)\phi]-\Bigl(K_{\rho s,m}-K_{\phi c,m}\Bigr)\right.\nonumber\\
& &\left.\times J_{m+1}(\kappa\rho_a^\prime\sin\theta)\sin[(m+1)\phi]\right\}\\
A_y&=&\pi a^2\mu_0G_0(R)\sum_{m=0}^\infty j^{m-1}\int_0^a\rho_a^\prime d\rho_a^\prime\nonumber\\
& & \times\left\{-\Bigl(K_{\rho c,m}-K_{\phi s,m}\Bigr)J_{m-1}(\kappa\rho_a^\prime\sin\theta)\sin[(m-1)\phi]-\Bigl(K_{\rho c,m}+K_{\phi s,m}\Bigr)\right.\nonumber\\
& &\left.\times J_{m+1}(\kappa\rho_a^\prime\sin\theta)\sin[(m+1)\phi]\right.\nonumber\\
& & \left. +\Bigl(K_{\rho s,m}+K_{\phi c,m}\Bigr)J_{m-1}(\kappa\rho_a^\prime\sin\theta)\cos[(m-1)\phi]+\Bigl(K_{\rho s,m}-K_{\phi c,m}\Bigr)\right.\nonumber\\
& &\left.\times J_{m+1}(\kappa\rho_a^\prime\sin\theta)\cos[(m+1)\phi]\right\}
\end{eqnarray}
In deriving Eq.(2.99)and (2.100), we use the formula of the integral representation of the Bessel function, given by
$$J_n(x)={j^{-n}\over 2\pi}\int_0^{2\pi}\exp(jx\cos\theta+jn\theta)d\theta={j^{-n}\over \pi}\int_0^\pi\exp(jx\cos\theta)\cos(n\theta)d\theta$$
By applying the relation
$$ A_\theta=A_x\cos\theta\cos\phi+A_y\cos\theta\sin\phi,\quad A_\phi=-A_x\sin\phi+A_y\cos\phi$$
the spherical coordinate components of the vector potential becomes
 \begin{eqnarray}
A_\theta&=&\pi a^2\mu_0G_0(R)\cos\theta\sum_{m=0}^\infty j^{m-1}\int_0^a \rho_a^\prime d\rho_a^\prime\nonumber\\
& &\times\left\{\Bigl[\Bigl(K_{\rho c,m}-K_{\phi s,m}\Bigr)J_{m-1}(\kappa\rho_a^\prime\sin\theta)-\Bigl(K_{\rho c,m}+K_{\phi s,m}\Bigr)J_{m+1}(\kappa\rho_a^\prime\sin\theta)\Bigr]\right.\nonumber\\
& &\left.\times\cos(m\phi)\right.\nonumber\\
& &\left. +\Big[\Bigr(K_{\rho s,m}+K_{\phi c,m}\Bigr)J_{m-1}(\kappa\rho_a^\prime\sin\theta)-\Bigl(K_{\rho s,m}-K_{\phi c,m}\Bigr)J_{m+1}(\kappa\rho_a^\prime\sin\theta)\Bigr]\right.\nonumber\\
& &\left.\times\sin(m\phi)\right\}\\
A_\phi&=&\pi a^2\mu_0G_0(R)\cos\theta\sum_{m=0}^\infty j^{m-1}\int_0^a\rho_a^\prime d\rho_a^\prime\nonumber\\
& &\times\left\{-\Bigl[\Bigl(K_{\rho c,m}-K_{\phi s,m}\Bigr)J_{m-1}(\kappa\rho_a^\prime\sin\theta)+\Bigl(K_{\rho c,m}+K_{\phi s,m}\Bigr)J_{m+1}(\kappa\rho_a^\prime\sin\theta)\Bigr]\right.\nonumber\\
& &\left.\times\sin(m\phi)\right.\nonumber\\
& &\left.+\Bigl[\Bigr(K_{\rho s,m}+K_{\phi c,m}\Bigr)J_{m-1}(\kappa\rho_a^\prime\sin\theta)+\Bigl(K_{\rho s,m}-K_{\phi c,m}\Bigr)J_{m+1}(\kappa\rho_a^\prime\sin\theta)\Bigr]\right.\nonumber\\
& &\left.\times\cos(m\phi)\right\}
\end{eqnarray}
where we have use the relation, given by
$$\cos(\alpha\pm\beta)=\cos\alpha\cos\beta\mp\sin\alpha\sin\beta$$
and
$$\sin(\alpha\pm\beta)=\sin\alpha\cos\beta\pm\cos\alpha\sin\beta$$
Substituting Eq.(2.58)$\sim$(2.61) into the Eq.(2.101)and (2.102), we have
\begin{eqnarray}
A_\theta&=&-j4\pi a^2\mu_0G_0(R)\cos\theta\sum_{m=0}^\infty B_{0n}J_{2n+{5\over 2}}(\kappa\sin\theta)(\kappa\sin\theta)^{-{3\over 2}}\nonumber+2\pi a^2\mu_0G_0(R)\cos\theta\nonumber\\
& &\times\sum_{m=1}^\infty j^{m-1}\sum_{n=0}^\infty\left\{A_{mn}\Bigl[J_{m+2n-{1\over 2}}(\kappa\sin\theta)+J_{m+2n+{3\over 2}}(\kappa\sin\theta)\Bigr](\kappa\sin\theta)^{-{1\over 2}}\right.\nonumber\\
& &\left. +B_{mn}\Bigl[J_{m+2n+{1\over 2}}(\kappa\sin\theta)-J_{m+2n+{5\over 2}}(\kappa\sin\theta)\Bigr](\kappa\sin\theta)^{-{3\over 2}}\right\}\cos(m\phi)\nonumber\\
& &+2\pi a^2 G_0(R)\cos\theta\nonumber\\
& &\times \sum_{m=1}^\infty j^{m-1}\sum_{n=0}^\infty\left\{C_{mn}\Bigl[J_{m+2n-{1\over 2}}(\kappa\sin\theta)+J_{m+2n+{3\over 2}}(\kappa\sin\theta)\Bigr](\kappa\sin\theta)^{-{1\over 2}}\right.\nonumber\\
& &\left.+D_{mn}[J_{m+2n+{1\over 2}}(\kappa\sin\theta)-J_{m+2n+{5\over 2}}(\kappa\sin\theta)\Bigr](\kappa\sin\theta)^{-{3\over 2}}\right\}\sin(m\phi)
\end{eqnarray}
\begin{eqnarray}
A_\phi&=&\pi a^2\mu_0G_0(R)\sum_{n=0}^\infty C_{0n}J_{2n+{3\over 2}}(\kappa\sin\theta)(\kappa\sin\theta)^{-{1\over 2}}+2\pi a^2G_0(R)\nonumber\\
& &\times\sum_{m=1}^\infty j^{m-1}\sum_{n=0}^\infty\left\{ A_{mn}\Bigl[J_{m+2n-{1\over 2}}(\kappa\sin\theta)-J_{m+2n+{3\over 2}}(\kappa\sin\theta)\Bigr](\kappa\sin\theta)^{-{1\over 2}}\right.\nonumber\\
& &\left. +B_{mn}\Bigl[J_{m+2n+{1\over 2}}(\kappa\sin\theta)+J_{m+2n+{5\over 2}}(\kappa\sin\theta)\Bigr](\kappa\sin\theta)^{-{3\over 2}}\right\}\sin(m\phi)\nonumber\\
& & +2\pi a^2G_0(R)\nonumber\\
& &\times\sum_{m=1}^\infty j^{m-1}\sum_{n=0}^\infty\left\{ C_{mn}\Bigl[J_{m+2n-{1\over 2}}(\kappa\sin\theta)-J_{m+2n+{3\over 2}}(\kappa\sin\theta)\Bigr](\kappa\sin\theta)^{-{1\over 2}}\right.\nonumber\\
& &\left. +D_{mn}\Bigl[J_{m+2n+{1\over 2}}(\kappa\sin\theta)+J_{m+2n+{5\over 2}}(\kappa\sin\theta)\Bigr](\kappa\sin\theta)^{-{3\over 2}}\right\}\cos(m\phi)
\end{eqnarray}
where we have again used the closure relation for the Hankel transform and the property of the delta function.
\paragraph{}Next we evaluate $A_z^d$ and $F_z^d$ given in Eq.(2.34) and (2.35), directly by applying the stationary  phase method of integration. These integral can be written in the form, given by
$$I_{nt}=\int_0^\infty \widetilde P(\xi) J_m(\rho_a\xi)\exp\Bigl[-\sqrt{\xi^2-\kappa^2}z_a\Bigr]\xi^{-1}d\xi$$
where we assume that $\widetilde P(\xi)$ is slowly varying function. To perform this integration asymptotically, we use the integral representation for the Bessel function given by
$$J_m(\rho_a\xi)={j^m\over 2\pi}\int_{-\pi}^{+\pi}\exp(-j\xi\rho_a\cos\alpha-jm\alpha)d\alpha$$
Now transforming the cylindrical coordinate variable $(\rho_a,z_a)$ into the polar coordinate variables $(R_a,\theta)$ through $\rho_a=R_a\sin\theta$ and $z_a=R_a\cos\theta$. And the integration variable $\xi$ can be change into $\beta$ by $\xi=\kappa\sin\beta$. Then the integral $I_{nt}$ changes into
\begin{eqnarray}
I_{nt}&=&{j^m\over 2\pi}\int_{-\pi}^{+\pi}\exp(-j\kappa R_a\sin\beta\sin\theta\cos\alpha-jm\alpha)d\alpha\nonumber\\
& &\times\int_C\widetilde P(\kappa\sin\beta)\exp[-j\kappa R_a\cos\beta\cos\theta]{\cos\beta\over\sin\beta}d\beta
\end{eqnarray}
where the contour $C$ is running along $(0,0)\to ({\pi\over 2},0)\to({\pi\over 2},\infty)$ in the complex $\beta$-plane. Stationary points are located at $(\alpha_0,\beta_0)$, which satisfied the equations
$$\sin\theta\cos\beta\cos\alpha_0-\cos\theta\sin\beta_0=0,\quad \kappa R_a\sin\theta\sin\beta_0\sin\alpha-m=0$$
when the value of $\kappa R_a$ is sufficiently large, $\alpha_0$ may be set to $0$, so that the approximate stationary points are given by
$$ \alpha_0=0,\quad \beta_0=\theta$$
Application of the standard process of the method yields the result,
\begin{eqnarray}
 I_{nt}=\exp\Bigl(j{m+1\over 2}\pi\Bigr){\exp(-j\kappa R_a)\over \kappa R_a}\widetilde P(\kappa\sin\theta){\cos\theta\over \sin^2\theta}
\end{eqnarray}
Now applying this formula to the vector potential given in Eq.(2.34) and (2.35), we have
\begin{eqnarray}
A_z^d(r)&=&j\mu_0 a^2{\exp(-j\kappa R)\over R}{\cos\theta\over\sin\theta}\sum_{n=0}^\infty B_{0n} J_{2n+{5\over 2}}(\kappa\sin\theta)(\kappa\sin\theta)^{-{3\over 2}}\nonumber\\
& &+\mu_0 a^2{\exp(-j\kappa R)\over 2R}{\cos\theta\over\sin\theta}\sum_{m=1}^\infty j^{m+1}\sum_{n=0}^\infty\left\{\Bigl[A_{mn}\Xi_{mn}^+(\kappa\sin\theta)-B_{mn}\Gamma_{mn}^-(\kappa\sin\theta)\Bigr]\cos(m\phi)\right.\nonumber\\
& &\left.+\Bigl[C_{mn}\Xi_{mn}^+(\kappa\sin\theta)+D_{mn}\Gamma_{mn}^-(\kappa\sin\theta)\Bigr]\sin(m\phi)\right\}\\
F_z^d(r)&=&j\epsilon_0 a^2Z_0{\exp(-j\kappa R)\over R}\sum_{n=0}^\infty C_{0n}J_{2n+{3\over 2}}(\kappa\sin\theta)(\kappa\sin\theta)^{-{1\over 2}}\nonumber\\
& &+\epsilon_0 a^2Z_0{\exp(-j\kappa R)\over 2R}\sum_{m=1}^\infty j^{m+1}\sum_{n=0}^\infty\left\{\Bigl[C_{mn}\Xi_{mn}^-(\kappa\sin\theta)+D_{mn}\Gamma_{mn}^+(\kappa\sin\theta)\Bigr]\cos m\phi\right.\nonumber\\
& &\left.+\Bigl[-A_{mn}\Xi_{mn}^-(\kappa\sin\theta)+B_{mn}\Gamma_{mn}^+(\kappa\sin\theta)\Bigr]\sin m\phi\right\}
\end{eqnarray}
\paragraph{}
In the far field only the $\theta$ and $\phi$ components of the $\mathbf{E}$ and $\mathbf{H}$ fields are dominant. Although the radial components are not necessarily zero, they are negligible compared to the $\theta$ and $\phi$ components. Thus for far field observation, we have
\begin{eqnarray}
E_\theta&=&-j\omega A_\theta=j\omega\sin\theta A_z\\
H_\theta&=&-j\omega F_\theta=j\omega\sin\theta F_z=-Y_0 E_\phi\\
A_\phi&=&Z_0\sin\theta F_z
\end{eqnarray}
or
\begin{eqnarray}
E_\theta&=&a{\exp(-jR)\over R}D_\theta(\theta,\phi)\\
E_\phi&=&a{\exp(-jR)\over R}D_\phi(\theta,\phi)
\end{eqnarray}
where
\begin{eqnarray} 
D_\theta(\theta,\phi)&=&-Z_0 \kappa a\cos\theta\sum_{n=0}^\infty B_{0n}J_{2n+{5\over 2}}(\kappa\sin\theta)(\kappa\sin\theta)^{-{3\over 2}}\nonumber\\
& &+j{Z_0\over 2}ka\cos\theta\sum_{m=1}^\infty j^{m+1}\sum_{n=0}^\infty\Bigl\{\Bigl[A_{mn}\Xi_{mn}^+(\kappa\sin\theta)-B_{mn}\Gamma_{mn}^-(\kappa\sin\theta)\Bigr]\cos m\phi\nonumber\\
& & +\Bigl[C_{mn}\Xi_{mn}^+(\kappa\sin\theta)+D_{mn}\Gamma_{mn}^-(\kappa\sin\theta)\Bigr]\sin m\phi\Bigr\}\\
D_\phi(\theta,\phi)&=&Z_0 \kappa a\sum_{n=0}^\infty C_{0n}J_{2n+{3\over 2}}(\kappa\sin\theta)(\kappa\sin\theta)^{-{1\over 2}}\nonumber\\
& & +j{Z_0\over 2}\kappa a\sum_{m=1}^\infty j^{m+1}\sum_{n=0}^\infty\Bigl\{\Bigl[C_{mn}\Xi_{mn}^-(\kappa\sin\theta)+D_{mn}\Gamma_{mn}^+(\kappa\sin\theta)\Bigr]\cos m\phi\nonumber\\
& &+\Bigl[-A_{mn}\Xi_{mn}^-(\kappa\sin\theta)+B_{mn}\Gamma_{mn}^+(\kappa\sin\theta)\Bigr]\sin m\phi\Bigr\}
\end{eqnarray}
These are the expression for the far-field pattern diffracted by a perfectly conducting circular disk. Also, if we use EQ.(2.109)$\sim$(2.111) we find that Eq.(2.107) and (2.108) agree completely with Eq.(2.106). Therefore we will used Eq.(2.106) directly to find the far-field pattern diffracted by a circular hole in perfectly conducting plate.

\section{The Expressions for the Fields Diffracted by a Circular Hole in a perfectly Conducting Plate}

\paragraph{}The diffraction of electromagnetic plane wave by a circular hole in a perfectly conducting plane is a complementary problem of the scattering by a disk and the solution is obtained directly by using the result of the disk problem via Babinets's principle.

\section{Electric Filed Distribution}

\paragraph{}
From Eq.(2.38)$\sim$(2.41), the expression for the electromagnetic filed can be express in matrix form, given by
\begin{eqnarray}
\left[\matrix{E_{\rho c,m}(\rho_a)\cr E_{\phi s,m}(\rho_a)}\right]&=&\int_0^\infty\Bigl[H^-(\xi\rho_a)\Bigr]\left[\matrix{j\sqrt{\xi^2-\kappa^2}\widetilde f_{cm}(\xi)\xi^{-1}\cr\widetilde g_{sm}(\xi)(x)^{-1}}\right]\xi d\xi\nonumber\\
& &=\int_0^\infty\Bigl[H^-(\xi\rho_a)\Bigr]\left[\matrix{\widetilde E_{\rho c,m}(\xi)\cr\widetilde E_{\phi s,m}(\xi)}\right]\xi d\xi\\
\left[\matrix{E_{\rho s,m}(\rho_a)\cr E_{\phi c,m}(\rho_a)}\right]&=&\int_0^\infty\Bigl[H^+(\xi\rho_a)\Bigr]\left[\matrix{j\sqrt{\xi^2-\kappa^2}\widetilde f_{sm}(\xi)\xi^{-1}\cr\widetilde g_{cm}(\xi)(\xi)^{-1}}\right]\xi d\xi\nonumber\\
& &=\int_0^\infty\Bigl[H^+(\xi\rho_a)\Bigr]\left[\matrix{\widetilde E_{\rho s,m}(\xi)\cr\widetilde E_{\phi c,m}(\xi)}\right]\xi d\xi\\
\left[\matrix{H_{\rho c,m}(\rho_a)\cr H_{\phi s,m}(\rho_a)}\right]&=&\pm Y_0\int_0^\infty\Bigl[H^-(\xi\rho_a)\Bigr]\left[\matrix{j\sqrt{\xi^2-\kappa^2}\widetilde g_{cm}(\xi)(\kappa\xi)^{-1}\cr -\kappa\widetilde f_{sm}(\xi)\xi^{-1}}\right]\xi d\xi\\
\left[\matrix{H_{\rho s,m}(\rho_a)\cr H_{\phi c,m}(\rho_a)}\right]&=&\pm Y_0\int_0^\infty\Bigl[H^+(\xi\rho_a)\Bigr]\left[\matrix{j\sqrt{\xi^2-\kappa^2}\widetilde g_{sm}(\xi)(\kappa\xi)^{-1}\cr -\kappa \widetilde f_{cm}(\xi)\xi^{-1}}\right]\xi d\xi
\end{eqnarray}
where the kernel matrices $[H^\pm(\xi\rho_a)]$ are defined by
$$\Bigl[H^\pm(\xi\rho_a)\Bigl]=\left[\matrix{J_m^\prime(\xi\rho_a)\qquad\pm{m\over \xi\rho_a}J_m(\xi\rho_a)\cr \pm{m\over \xi\rho_a}J_m(\xi\rho_a)\quad J_m^\prime(\xi\rho_a)}\right]$$
Using the boundary condition, the dual integral in this case can be written as
\begin{eqnarray}
\int_0^\infty\Bigl[H^-(\xi\rho_a)\Bigr]\left[\matrix{\widetilde E_{\rho c,m}(\xi)\cr \widetilde E_{\phi s,m}(\xi)}\right]\xi d\xi=0\quad \rho_a\geq 1\\
\int_0^\infty\Big[H^+(\xi\rho_a)\Bigr]\left[\matrix{\widetilde E_{\rho s,m}(\xi)\cr\widetilde E_{\phi c,m}(\xi)}\right]\xi d\xi=0\quad \rho_a\geq 1
\end{eqnarray}
\begin{eqnarray}
Y_0\int_0^\infty\Bigl[H^-(\xi\rho_a)\Bigr]\left[\matrix{j\sqrt{\xi^2-\kappa^2}\widetilde g_{cm}(\xi)(\kappa\xi)^{-1}\cr -\kappa\widetilde f_{sm}(\xi)\xi^{-1}}\right]\xi d\xi+\left[\matrix{H^i_{\rho c,m}(\rho_a)\cr H^i_{\phi s,m}(\rho_a)}\right]=0\quad\rho_a\leq1\\
Y_0\int_0^\infty\Bigl[H^+(\xi\rho_a)\Bigr]\left[\matrix{j\sqrt{\xi^2-\kappa^2}\widetilde g_{sm}(\xi)(\kappa\xi)^{-1}\cr -\kappa\widetilde f_{cm}(\xi)\xi^{-1}}\right]\xi d\xi+\left[\matrix{H^i_{\rho s,m}(\rho_a)\cr H^i_{\phi c,m}(\rho_a)}\right]=0\quad\rho_a\leq1
\end{eqnarray}
where $H^i_{\phi c,m}$ and $H^i_{\rho s,m}$ represent the $\cos m\phi$ and $\sin m\phi$ parts of the incident wave $H^i_\rho$, respectively, and the same is true for $H^i_{\phi c,m}$ and $H^i_{\phi s,m}$. The expression for these factors are given by
\begin{eqnarray}
H^i_{\rho c}(\rho_a)&=&-jY_0E_1\cos\theta_0\epsilon_m j^m J_m^\prime(\kappa\rho_a\sin\theta_0)\\
H^i_{\rho s}(\rho_a)&=&jY_0E_2\epsilon_m j^m {m\over \kappa\rho_a\sin\theta_0}J_m(\kappa\rho_a\sin\theta_0)\\
H^i_{\phi c}(\rho_a)&=&jY_0E_2\epsilon_m j^m J_m^\prime(\kappa\rho_a\sin\theta_0)\\
H^i_{\phi s}(\rho_a)&=&jY_0E_1\cos\theta_0\epsilon_m j^m{m\over \kappa\rho_a\sin\theta_0}J_m(\kappa\rho_a\sin\theta_0)
\end{eqnarray}
\indent The aperture electric field can be expanded in a manner similar to the disk problem. It is noted that $(E_\rho,E_\phi)$ satisfy the vector Helmholtz equation $\nabla^2 \textbf{E}+\kappa^2 \textbf{E}=0$ in circular cylindrical coordinates. Furthermore, $(E_\rho, E_\phi)$ have the property $E_\rho\sim(1-\rho_a^2)^{-{1\over2}}$ and $E_\phi\sim(1-\rho_a^2)^{{1\over2}}$ near the edge of the hole. By taking into these facts and using the property of the Weber-Schafheitlin's discontinuous integral, we set
\begin{eqnarray}
E_{\rho c,m}(\rho_a)&=&\sum_{n=0}^\infty\Bigl[A_{mn}F^+_{mn}(\rho_a)-B_{mn}G^-_{mn}(\rho_a)\Bigr]\\
E_{\rho s,m}(\rho_a)&=&\sum_{n=0}^\infty\Bigl[C_{mn}F^+_{mn}(\rho_a)+D_{mn}G^-_{mn}(\rho_a)\Bigr]\\
E_{\phi s,m}(\rho_a)&=&\sum_{n=0}^\infty\Bigl[-A_{mn}F^-_{mn}(\rho_a)+B_{mn}G^+_{mn}(\rho_a)\Bigr]\\
E_{\phi c,m}(\rho_a)&=&\sum_{n=0}^\infty\Bigl[C_{mn}F^-_{mn}(\rho_a)+D_{mn}G^+_{mn}(\rho_a)\Bigr]
\end{eqnarray}
where
\begin{eqnarray}
F^\pm_{mn}(\rho_a)&=&\int_0^\infty\Bigl[J_{|m-1|}(\eta\rho_a)J_{|m-1|+2n+{1\over2}}(\eta)\pm J_{m+1}(\eta\rho_a)J_{m+2n+{3\over 2}}\eta\Bigr]\eta^{{1\over 2}}d\eta\\
G^\pm_{mn}(\rho_a)&=&\int_0^\infty\Bigl[J_{|m-1|}(\eta\rho_a)J_{|m-1|+2n+{3\over 2}}(\eta)\pm J_{m+1}(\eta\rho_a)J_{m+2n+{5\over 2}}\eta\Bigr]\eta^{-{1\over 2}}d\eta
\end{eqnarray}
where $A_{mn}\sim D_{mn}$ are the expansion coefficients and are to be determined from the remaining boundary condition that the tangential components of the magnetic field are continuous on the aperture. The corresponding spectrum function can be derived by applying the vector Hankel transform, given by
\begin{eqnarray}
\left[\matrix{\widetilde E_{\rho c,m}(\xi)\cr\widetilde E_{\phi s,m}(\xi)}\right]&=&\int_0^\infty\Bigl[H^-(\xi\rho_a)\Bigr]\left[\matrix{\widetilde E_{\rho c,m}(\rho_a)\cr \widetilde E_{\phi s,m}(\rho_a)}\right]\rho_a d\rho_a\\
\left[\matrix{\widetilde E_{\rho s,m}(\xi) \cr\widetilde E_{\phi c,m}(\xi)}\right]&=&\int_0^\infty \Bigl[H^+(\xi\rho_a)\Bigr]\left[\matrix{\widetilde E_{\rho s,m}(\rho_a)\cr\widetilde E_{\phi c,m}(\rho_a)}\right]\rho_a d\rho_a
\end{eqnarray}
\indent Substituting Eq.(2.128)$\sim$(2.131) into Eq.(2.134) and (2.135) and perform the integration, then the spectrum function for the current density can be determined. The result is given by
\begin{eqnarray}
\widetilde E_{\rho c,m}(\xi)&=&\sum_{n=0}^\infty\Bigl[A_{mn}\Xi^-_{mn}(\xi)-B_{mn}\Gamma^+_{mn}(\xi)\Bigr]\\
\widetilde E_{\phi s,m}(\xi)&=&\sum_{n=0}^\infty\Bigl[-A_{mn}\Xi^+_{mn}(\xi)+B_{mn}\Gamma^-_{mn}(\xi)\Bigr]\\
\widetilde E_{\rho s,m}(\xi)&=&\sum_{n=0}^\infty\Bigl[C_{mn}\Xi^-_{mn}(\xi)+D_{mn}\Gamma^+_{mn}(\xi)\Bigr]\\
\widetilde E_{\phi c,m}(\xi)&=&\sum_{n=0}^\infty\Bigl[C_{mn}\Xi^+_{mn}(\xi)+D_{mn}\Gamma^-_{mn}(\xi)\Bigr]
\end{eqnarray}
for $m\geq1$ and
\begin{eqnarray}
\widetilde E_{\rho c,0}(\xi)=-2\sum_{n=0}^\infty A_{0n}J_{2n+{3\over 2}}(\xi)\xi^{-{1\over 2}}\qquad\widetilde E_{\phi s,0}(\xi)=-2\sum_{n=0}^\infty B_{0n}J_{2n+{5\over 2}}(\xi)\xi^{-{3\over 2}}\\
\widetilde E_{\rho s,0}(\xi)=-2\sum_{n=0}^\infty C_{0n}J_{2n+{3\over 2}}(\xi)\xi^{-{1\over 2}}\qquad\widetilde E_{\phi c,0}(\xi)=-2\sum_{n=0}^\infty D_{0n}J_{2n+{5\over 2}}(\xi)\xi^{-{3\over 2}}
\end{eqnarray}
for $m=0$. From Eq.(2.116) and (2.117) we can express the spectral functions $\widetilde f_{cm}(\xi)\sim\widetilde g_{sm}(\xi)$ of the vector potentials in terms of those of the aperture distribution. This mean that the surface magnetic field can be expressed in term of the spectrum functions of the surface electric field.

\section{Derivation of the Expansion Coefficients}

\paragraph{}
From Eq.(2.116) and (2.117), the spectrum function $\widetilde f_{cm}(\xi)\sim\widetilde g_{mn}(\xi)$ may be expressed in term of $\widetilde E(\xi)$, that is
\begin{eqnarray}
\widetilde f_{cm}(\xi)\xi^{-1}={1\over j\sqrt{\xi^2-\kappa^2}}\widetilde E_{\rho c,m}(\xi)\qquad\widetilde g_{sm}(\xi)\xi^{-1}=\widetilde E_{\phi s,m}(\xi)\\
\widetilde f_{sm}(\xi)\xi^{-1}={1\over j\sqrt{\xi^2-\kappa^2}}\widetilde E_{\rho s,m}(\xi)\qquad\widetilde g_{cm}(\xi)\xi^{-1}=\widetilde E_{\phi c,m}(\xi)
\end{eqnarray}
\indent Substituting the above relations into Eq.(2.122) and (2.123), the continuities of the tangential components of the magnetic field are written as follows
\begin{eqnarray}
Y_0\int_0^\infty\Bigl[H^-(\xi\rho_a)\Bigr]\left[\matrix{{j\sqrt{\xi^2-\kappa^2}\over\kappa}\widetilde E_{\phi c,m}(\xi)\cr j{\kappa\over \sqrt{\xi^2-\kappa^2}}\widetilde E_{\rho s,m}(\xi)}\right]\xi d\xi+\left[\matrix{H^i_{\rho c,m}(\rho_a)\cr H^i_{\phi s,m}(\rho_a)}\right]=0\quad \rho_a\leq1
\end{eqnarray}
\begin{eqnarray}
Y_0\int_0^\infty\Bigl[H^+(\xi\rho_a)\Bigr]\left[\matrix{{j\sqrt{\xi^2-\kappa^2}\over \kappa}\widetilde E_{\phi s,m}(\xi)\cr j{\kappa\over \sqrt{\xi^2-\kappa^2}}\widetilde E_{\phi c,m}(\xi)}\right]\xi d\xi+\left[\matrix{H^i_{\rho s,m}(\rho_a)\cr H^i_{\phi c,m}(\rho_a)}\right]=0\quad \rho_a\leq 1
\end{eqnarray}
\indent Eq.(2.144) and (2.145) are projected into function space with element $v^m_n(\rho_a^2)$ for $H_\rho$ and $u^m_n(\rho_a^2)$ for $H_\phi$, then the matrix equation for the expansion coefficients $A_{mn}\sim D_{mn}$. The result are given by
\begin{eqnarray}
j\sum_{n=0}^\infty C_{mn}\left\{{1\over \kappa}\int_0^\infty {\sqrt{\xi^2-\kappa^2}\over \xi}\Bigl[J_{m+2n-{1\over 2}}(\xi)+J_{m+2n+{3\over 2}}(\xi)\Bigr]\Bigl[\alpha^m_p J_{m+2p+{1\over 2}}(\xi)-(\alpha^m_p+3) J_{m+2p+{5\over 2}}(\xi)\Bigr]d\xi\right.\nonumber\\
\left.-\kappa m\int_0^\infty{1\over\xi\sqrt{\xi^2-\kappa^2}}\Bigl[J_{m+2n-{1\over 2}}(\xi)-J_{m+2n+{3\over 2}}(\xi)\Bigr]\Bigl[J_{m+2p+{1\over 2}}(\xi)+J_{m+2p+{5\over 2}}(\xi)\Bigr]d\xi\right\}\nonumber\\
+j\sum_{n=0}^\infty D_{mn}\left\{{1\over\kappa}\int_0^\infty{\sqrt{\xi^2-\kappa^2}\over \xi^2}\Bigl[J_{m+2n+{1\over 2}}(\xi)-J_{m+2n+{5\over 2}}(\xi)\Bigr]\Bigl[\alpha^m_n J_{m+2p+{1\over 2}}(\xi)-(\alpha^m_p+3)J_{m+2p+{5\over 2}}(\xi)\Bigr]d\xi\right.\nonumber\\
\left.-\kappa m\int_0^\infty {1\over \xi^2\sqrt{\xi^2-\kappa^2}}\Bigl[J_{m+2n+{1\over 2}}(\xi)+J_{m+2n+{5\over 2}}(\xi)\Bigr]\Bigl[J_{m+2p+{1\over 2}}(\xi)+J_{m+2p+{5\over 2}}(\xi)\Bigr]d\xi\right\}\nonumber\\
=j2E_1\cos\theta_0 j^m\Bigl[\alpha^m_p J_{m+2p+{1\over 2}}(\kappa\sin\theta_0)-(\alpha^m_p+3)J_{m+2p+{5\over 2}}(\kappa\sin\theta_0)\Bigr](\kappa\sin\theta_0)^{-{3\over 2}}\nonumber\\
\qquad m=1,2,3,\cdots ,\quad p=0,1,2,3,\cdots ,\nonumber\\
\end{eqnarray}
\begin{eqnarray}
-j\sum_{n=0}^\infty C_{mn}\left\{{m\over \kappa}\int_0^\infty \sqrt{\xi^2-\kappa^2}\Bigl[J_{m+2n-{1\over 2}}(\xi)+J_{m+2n+{3\over 2}}(\xi)\Bigr]\Bigl[J_{m+2p-{1\over 2}}(\xi)+J_{m+2p+{3\over 2}}(\xi)\Bigr]d\xi\right.\nonumber\\
\left.-\kappa\int_0^\infty{1\over\sqrt{\xi^2-\kappa^2}}\Bigl[J_{m+2n-{1\over 2}}(\xi)+J_{m+2n+{3\over 2}}(\xi)\Bigr]\Bigl[\alpha^m_p J_{m+2p-{1\over 2}}(\xi)-(\alpha^m_p+1) J_{m+2p+{5\over 2}}(\xi)\Bigr]d\xi\right\}\nonumber\\
-j\sum_{n=0}^\infty D_{mn}\left\{{m\over \kappa}\int_0^\infty{\sqrt{\xi^2-\kappa^2}\over \xi}\Bigl[J_{m+2n+{1\over 2}}(\xi)-J_{m+2n+{5\over 2}}(\xi)\Bigr]\Bigl[J_{m+2p+{1\over 2}}(\xi)+J_{m+2p+{3\over 2}}(\xi)\Bigr]d\xi\right.\nonumber\\
\left.-\kappa\int_0^\infty {1\over\xi\sqrt{\xi^2-\kappa^2}}\Bigl[J_{m+2n+{1\over 2}}(\xi)+J_{m+2n+{5\over 2}}(\xi)\Bigr]\Bigl[\alpha^m_p J_{m+2p-{1\over 2}}(\xi)-(\alpha^m_p+1)J_{m+2p+{3\over 2}}(\xi)\Bigr]d\xi\right\}\nonumber\\
=-j2E_1\cos\theta_0 j^mm\Bigl[J_{m+2p-{1\over 2}}(\kappa\sin\theta_0)+J_{m+2p+{3\over 2}}(\kappa\sin\theta_0)\Bigr](\kappa\sin\theta_0)^{-{1\over 2}}\nonumber\\
\qquad m=1,2,3,\cdots ,\quad p=0,1,2,3,\cdots ,\nonumber\\
\end{eqnarray}
\begin{eqnarray}
-j\sum_{n=0}^\infty A_{mn}\left\{{1\over \kappa}\int_0^\infty{\sqrt{\xi^2-\kappa^2}\over \xi}\Bigl[J_{m+2n-{1\over 2}}(\xi)+J_{m+2n+{3\over 2}}(\xi)\Bigr]\Bigl[\alpha^m_p J_{m+2p+{1\over 2}}(\xi)-(\alpha^m_p+3)J_{m+2p+{5\over 2}}(\xi)\Bigr]d\xi\right.\nonumber\\
\left.-\kappa m\int_0^\infty {1\over \xi\sqrt{\xi^2-\kappa^2}}\Bigl[J_{m+2n-{1\over 2}}(\xi)-J_{m+2n+{3\over 2}}(\xi)\Bigr]\Bigl[J_{m+2p+{1\over 2}}(\xi)+J_{m+2p+{5\over 2}}(\xi)\Bigr]d\xi\right\}\nonumber\\
+j\sum_{n=0}^\infty B_{mn}\left\{{1\over \kappa}\int_0^\infty{\sqrt{\xi^2-\kappa^2}\over \xi^2}\Bigl[J_{m+2n+{1\over 2}}(\xi)-J_{m+2n+{5\over 2}}(\xi)\Bigr]\Bigl[\alpha^m_p J_{m+2p+{1\over 2}}(\xi)-(\alpha^m_p+3)J_{m+2p+{5\over 2}}(\xi)\Bigr]d\xi\right.\nonumber\\
\left.-\kappa m\int_0^\infty {1\over \xi^2\sqrt{\xi^2-\kappa^2}}\Bigl[J_{m+2n+{1\over 2}}(\xi)+J_{m+2n+{5\over 2}}(\xi)\Bigr]\Bigl[J_{m+2p+{1\over 2}}(\xi)+J_{m+2p+{5\over 2}}(\xi)\Bigr]d\xi\right\}\nonumber\\
=-j2E_2j^mm\Bigl[J_{m+2p+{1\over 2}}(\kappa\sin\theta_0)+J_{m+2p+{5\over 2}}(\kappa\sin\theta_0)\Bigr](\kappa\sin\theta_0)^{-{3\over 2}}\nonumber\\\qquad m=1,2,3,\cdots ,\quad p=0,1,2,3,\cdots ,\nonumber\\
\end{eqnarray}
\begin{eqnarray}
-j\sum_{n=0}^\infty A_{mn}\left\{{m\over \kappa}\int_0^\infty \sqrt{\xi^2-\kappa^2}\Bigl[J_{m+2n-{1\over 2}}(\xi)+J_{m+2n+{3\over 2}}(\xi)\Bigr]\Bigl[J_{m+2p-{1\over 2}}(\xi)+J_{m+2p+{3\over 2}}(\xi)\Bigr]d\xi\right.\nonumber\\
\left.-\kappa\int_0^\infty {1\over\sqrt{\xi^2-\kappa^2}}\Bigl[J_{m+2n-{1\over 2}}(\xi)-J_{m+2n+{3\over 2}}(\xi)\Bigr]\Bigl[\alpha^m_pJ_{m+2p-{1\over 2}}(\xi)-(\alpha^m_p+1)J_{m+2p+{3\over 2}}(\xi)\Bigr]d\xi\right\}\nonumber\\
+j\sum_{n=0}^\infty B_{mn}\left\{{m\over \kappa}\int_0^\infty{\sqrt{\xi^2-\kappa^2}\over \xi}\Bigl[J_{m+2n+{1\over 2}}(\xi)-J_{m+2n+{5\over 2}}(\xi)\Bigr]\Bigl[J_{m+2p-{1\over 2}}(\xi)+J_{m+2p+{3\over 2}}(\xi)\Bigr]d\xi\right.\nonumber\\
\left.-\kappa\int_0^\infty{1\over \xi\sqrt{\xi^2-\kappa^2}}\Bigl[J_{m+2n+{1\over 2}}(\xi)+J_{m+2n+{5\over 2}}(\xi)\Bigr]\Bigl[\alpha^m_p J_{m+2p-{1\over 2}}(\xi)-(\alpha^m_p+1)J_{m+2p+{3\over 2}}(\xi)\Bigr]d\xi\right\}\nonumber\\
=-j2E_2j^m\Bigl[\alpha^m_pJ_{m+2p-{1\over 2}}(\kappa\sin\theta_0)-(\alpha^m_p+1)J_{m+2p+{3\over 2}}(\kappa\sin\theta_0)\Bigr](\kappa\sin\theta_0)^{-{1\over 2}}\nonumber\\
\qquad m=1,2,3,\cdots ,\quad p=0,1,2,3,\cdots ,\nonumber\\
\end{eqnarray}

\begin{eqnarray*}
\sum_{n=0}^\infty D_{0n}{1\over\kappa}\int_0^\infty{\sqrt{\xi^2-\kappa^2}\over \xi^2}J_{2n+{5\over 2}}(\xi)\Bigl[-pJ_{2p+{1\over 2}}(\xi)+(p+1.5)J_{2p+{5\over 2}}(\xi)\Bigr]d\xi\nonumber\\
=-E_1\cos\theta_0\Bigl[-pJ_{2p+{1\over 2}}(\kappa\sin\theta_0)+(p+1.5)J_{2p+{5\over 2}}(\kappa\sin\theta_0)\Bigr](\kappa\sin\theta_0)^{-{3\over 2}}\nonumber\\
\end{eqnarray*}

\begin{eqnarray}
\sum_{n=0}^\infty A_{0n}\kappa\int_0^\infty{1\over\sqrt{\xi^2-\kappa^2}}J_{2n+{3\over 2}}(\xi)\Bigl[-pJ_{2p-{1\over 2}}(\xi)+(p+0.5)J_{2p+{3\over 2}}(\xi)\Bigr]d\xi\nonumber\\
=-E_2\Bigl[-pJ_{2p-{1\over 2}}(\kappa\sin\theta_0)+(p+0.5)J_{2p+{3\over 2}}(\kappa\sin\theta_0)\Bigr](\kappa\sin\theta_0)^{-{1\over 2}}\nonumber\\
\qquad p=0,1,2,3,\cdots ,\nonumber\\
\end{eqnarray}

\section{Far Field Expression}

\paragraph{}
Far filed is obtained by applying the formula given in Eq.(2.106), to the expression of the vector potential, the result is given by
\begin{eqnarray}
A^d_z(\rho_a,\phi,z_a)&=&j2\mu_0a^2Y_0{\exp(-j\kappa R)\over R}{1\over \sin\theta}\sum_{n=0}^\infty A_{0n}J_{2n+{3\over 2}}(\kappa\sin\theta)(\kappa\sin\theta)^{-{1\over 2}}\nonumber\\
& &+\mu_0a^2Y_0{\exp(-j\kappa R)\over R}\sum_{m=1}^\infty j^{m-1}\sum_{n=0}^\infty\left\{\Bigl[A_{mn}\Xi^-_{mn}(\kappa\sin\theta)-B_{mn}\Gamma^+_{mn}(\kappa\sin\theta)\Bigr]\cos m\phi\right.\nonumber\\
& &\left.+\Bigl[C_{mn}\Xi^-_{mn}(\kappa\sin\theta)+D_{mn}\Gamma^+_{mn}(\kappa\sin\theta)\Bigr]\sin m\phi\right\}{1\over \sin\theta}\\
F^d_z(\rho_a,\phi,z_a)&=&-j2\epsilon_0 a^2{\exp(-j\kappa R)\over R}{\cos\theta\over\sin\theta}\sum_{n=0}^\infty D_{0n}J_{2n+{5\over 2}}(\kappa\sin\theta)(\kappa\sin\theta)^{-{3\over 2}}\nonumber\\
& &+\epsilon_0a^2{\exp(-j\kappa R)\over R}\sum_{m=1}^\infty j^{m-1}\sum_{n=0}^\infty\left\{\Bigl[C_{mn}\Xi^+_{mn}(\kappa\sin\theta)+D_{mn}\Gamma^-_{mn}(\kappa\sin\theta)\Bigr]\cos m\phi\right.\nonumber\\
& &\left.+\Bigl[-A_{mn}\Xi^+_{mn}(\kappa\sin\theta)+B_{mn}\Gamma^-_{mn}(\kappa\sin\theta)\Bigr]\sin m\phi\right\}{\cos\theta\over\sin\theta}\end{eqnarray}
In the far region we have again the relation given by
\begin{eqnarray}
E_\theta&=&-j\omega A_\theta=j\omega\sin\theta A_z\\
H_\theta&=&-j\omega F_\theta=-Y_0E_\phi\\
A_\phi&=&Z_0\sin\theta F_z
\end{eqnarray}
These are the expression for the far-field pattern diffracted by a circular hole in a perfectly conducting plate, when the observation point lie in the far region.
\pagebreak

\chapter{Numerical Computation}
\paragraph{} The numerical computation has been carried out for the far-filed pattern diffracted by a perfectly conducting disk for normal incidence. The results are then compared with the physical optics (PO) approximation solution. The main problem in computational work is the numerical solution of the infinite integrals to obtained the expansion coefficient. Here we have transform these integrals into infinite series which are more convenient for numerical computation. Moreover, the numerical results for these integrals can be also obtained directly by using the numerical integration via the Gauss-Legendre quadrature, which is a simple check for the validity of these series solution.

\section{Physical Optics Approximation Solution}

\paragraph{} We derive the approximate solution by using the physical optics to compare it with the results computed by the present method. We consider here the scattering by disk illuminated by a plane wave. The induced current is given by
\begin{eqnarray}
J&=&2n\times H^i=2Y_0\Bigl[E_2i_x+E_1\cos\theta_0 i_y\Bigr]\exp(j\kappa x\sin\theta_0)
\end{eqnarray}
The field produced by this current in the far region is derived from Eq.(2.87) as follows.
\begin{eqnarray}
A_x&=&2Y_0E_2\mu_0 G_0(R)\int_S \exp\Bigl[j\kappa\rho_a^\prime\sin\theta_0\cos\phi^\prime+j\kappa\rho_a^\prime\sin\theta\cos(\phi-\phi^\prime)\Bigr]dS\nonumber\\
& &=2Y_0E_2a^2\mu_0 G_0(R)\int_0^1\int_0^{2\phi}\exp\Bigl[j\kappa\rho_a^\prime\Theta\cos(\phi^\prime-\Phi)\Bigr]\rho_a^\prime d\rho_a^\prime d\phi^\prime\nonumber\\
& &=4\pi Y_0 e_2 a^2\mu_0 G_0(R)\int_0^1 J_0(\kappa\rho_a^\prime\Theta)\rho_a^\prime d\rho_a^\prime d\phi^\prime
\end{eqnarray}
where we set
\begin{eqnarray}
\sin\theta_0+\sin\theta\cos\phi&=&\Theta\cos\Phi,\qquad \sin\theta\sin\phi=\Theta\sin\Phi\nonumber\\
\Theta&=&\sqrt{(\sin\theta_0+\sin\theta\cos\phi)^2+(\sin\theta\sin\phi)^2}
\end{eqnarray}
The above integral can be readily carried out and we have
\begin{eqnarray}
A_x&=&2\pi Y_0E_2 a^2\mu_0 G_0(R){2J_1(\kappa\Theta)\over \kappa\Theta}
\end{eqnarray}
Similarly we have
\begin{eqnarray}
A_y&=&2\pi Y_0E_1a^2\cos\theta_0\mu_0 G_0(R){2J_1(\kappa\Theta)\over \kappa\Theta}
\end{eqnarray}
Hence from Eq.(2.109), we have
\begin{eqnarray}
E_\theta&\sim &-j\omega A_\theta=-2\pi a\kappa\cos\theta G_0(R)\Bigl(E_2\cos\phi+E_1\cos\theta_0\sin\phi\Bigr){2J_1(\kappa\Theta)\over \kappa\Theta}\\
E_\phi&\sim&-j\omega A_\phi=-j2\pi a\kappa G_0(R)\Bigl(-E_2\sin\phi+E_1\cos\theta_0\cos\phi\Bigr){2J_1(\kappa\Theta)\over \kappa\Theta}
\end{eqnarray}
These are the expression for the electric field, which is obtained by using the physical optics method and can be used for the computational purpose, to compare the result with those obtained by the Kobayashi potential method.
\section{Series Solution of the Integral $G(\alpha,\beta;\kappa)$}
Consider an evaluation of the integral defined by
\begin{eqnarray}
G(\alpha,\beta;\kappa)&=&\int_0^\infty{J_\alpha(\xi)J_\beta(\xi)\over\sqrt{\xi^2-\kappa^2}}d\xi
\end{eqnarray}
This integral was first evaluated by Nomura and Katsura. Here we have derive the series solution of this integral in a different way. That is, we define Eq.(3.8) as the limiting value of the integral given by
\begin{eqnarray}
G(\alpha,\beta;\kappa)&=&\lim_{v\rightarrow 0}\int_0^\infty {J_\alpha(\xi)J_\beta(\xi)\over\sqrt{\xi^2-\kappa^2}}\exp[-\sqrt{\xi^2-\kappa^2}v]d\xi
\end{eqnarray}
In the first step, we drive the series representation for larger value of $v(v>1)$, then the expressions converted into contour integral and derive the result which is valid for smaller value of $v(v<1)$. Using the integral representation for the product of the Bessel function given by
\begin{eqnarray}
J_\mu(\xi)J_\nu(\xi)&=&{2\over\pi}\int_0^{\pi\over 2}J_{\mu+\nu}(2\xi\cos\theta)\cos([\mu-\nu]\theta)d\theta
\end{eqnarray}
And other integral formula for the Bessel functions given by
\begin{eqnarray}
J_{\alpha+\beta}(2\xi\cos\theta)={1\over 2\pi j}\int_{-j\infty-\xi}^{+j\infty-\xi}{\Gamma(-t)\over\Gamma(\alpha+\beta+t+1)}(\xi\cos\theta)^{\alpha+\beta+2t}dt\quad(\xi<0)
\end{eqnarray}
Eq.(3.8) is transformed into
\begin{eqnarray}
G(\alpha,\beta;\kappa)&=&{1\over\pi^2 j}\int_{-j\infty-\xi}^{+j\infty-\xi}{\Gamma(-t)dt\over \Gamma(\alpha+\beta+t+1)}\int_0^{\pi\over 2}\cos(\theta)^{\alpha+\beta+2t}\cos[(\alpha-\beta)\theta]d\theta\nonumber\\
& &\times \int_0^\infty{1\over\sqrt{\xi^2-\kappa^2}}\xi^{\alpha+\beta+2t}\exp[-\sqrt{\xi^2-\kappa^2}v]d\xi
\end{eqnarray}
The integral with respect to $\theta$ and $\xi$ may be carried out with the results
\begin{eqnarray}
K_1&=&\int_0^{\pi\over 2}(\cos\theta)^{\alpha+\beta+2t}\cos[(\alpha-\beta)\theta]d\theta\nonumber\\
& &={\pi\Gamma(\alpha+\beta+2t+1)\over 2^{\alpha+\beta+2t+1}\Gamma(\alpha+t+1)\Gamma(\beta+t+1)}\\
K_2&=&\int_0^\infty{({\xi\over2})^{\alpha+\beta+2t}\over\sqrt{\xi^2-\kappa^2}}\exp[-\sqrt{\xi^2-\kappa^2}v]d\xi\nonumber\\
& &={\sqrt{\pi}\over2}\Gamma\Bigl[{1\over 2}(\alpha+\beta)+t+{1\over 2}\Bigr]{({\kappa\over 2})^{\alpha+\beta+2t}\over({\kappa v\over 2})^{{(\alpha+\beta)\over 2+t}}}[-Y_{{1\over 2}(\alpha+\beta)+t}(\kappa v)-jJ_{{1\over2}(\alpha+\beta)}(\kappa v)]
\end{eqnarray}
Eq.(3.14) is valid for $v>1$. Substituting these results into Eq.(3.12) we have
\begin{eqnarray}
G(\alpha,\beta;\kappa)&=&{1\over 4\sqrt{pi}j}\int_{-j\infty-\xi}^{+j\infty+\xi}{\Gamma(-t)\over \Gamma(\alpha+\beta+t+1)}{\Gamma(\alpha+\beta+2t+1)\Gamma[{1\over2}(\alpha+\beta)+t+{1\over 2}]\over \Gamma(\alpha+t+1)\Gamma(\beta+t+1)}\nonumber\\
& & \times {({\kappa\over 2})^{\alpha+\beta+2t}\over({\kappa v\over 2})^{(\alpha+\beta)\over 2+t}}[-Y_{{1\over 2}(\alpha+\beta)+t}(\kappa v)-jJ_{{1\over 2}(\alpha+\beta)+t}(\kappa v)]dt
\end{eqnarray}
The Neumann function in the above equation is replace by the relation used as its definition, which is given by
\begin{eqnarray}
Y_{{1\over 2}(\alpha+\beta)+t}(\kappa v)&=&{1\over \sin\{[{1\over 2}(\alpha+\beta)+t]\pi\}}\nonumber\\
& &\times\Bigl[\cos\Bigl\{({1\over 2}(\alpha+\beta)+t)\pi\Bigr\}J_{{1\over 2}(\alpha+\beta)+t}(\kappa v)-J_{-[{1\over 2}(\alpha+\beta)+t]}(\kappa v)\Bigr]
\end{eqnarray}
Using the above relation, $G(\alpha,\beta;\kappa)$ can be split into three parts, given by
\begin{eqnarray}
I_1&=&{1\over 4\sqrt{\pi}j}\int_{-j\infty-\xi}^{+j\infty-\xi}Q(t)\csc\Bigl[\Bigl({\alpha+\beta\over 2}+t\Bigr)\pi\Bigr]J_{-[{1\over2}(\alpha+\beta)+t]}(\kappa v)dt\\
I_2&=&-{1\over 4\sqrt{\pi}}\int_{-j\infty-\xi}^{+j\infty-\xi}Q(t)J_{{1\over 2}(\alpha+\beta)+t}(\kappa v)dt\\
I_3&=&-{1\over 4\sqrt{\pi}j}\int_{-j\infty-\xi}^{+j\infty-\xi}Q(t)\cot\Bigl[\Bigl({\alpha+\beta\over 2}+t\Bigr)\pi\Bigr]J_{{1\over 2}(\alpha+\beta)+t}(\kappa v)dt\\
Q(t)&=&{\Gamma(-t)\Gamma(\alpha+\beta+2t+1)\Gamma[{1\over 2}(\alpha+\beta)+t+{1\over 2}]\over \Gamma(\alpha+\beta+t+1)\Gamma(\alpha+t+1)\Gamma(\beta+t+1)}({\kappa\over 2v})^{{1\over 2}(\alpha+\beta)+t}
\end{eqnarray}
\indent In the limit $|t|\to \infty$, the integral of $I_1$ approaches to $\exp[2t\ln(2/v)]$, while those of $I_2$ and $I_3$ approach to $\exp[-t\ln n]$. Therefore, we may close the contour of $I_1$ in the left half plane for $v < 2$, and those of $I_2$ and $I_3$ in the right half plane. Since the evaluation of the integral depends on the indices $\alpha$ and $\beta$, we consider the following cases.\\
\smallskip
\textbf{(a) Evaluation of the Integral $I_1$}\\
\smallskip
The integrand of the integral $I_1$ has the singularities\\
(a) simple pole at $t=-\ell \quad \Bigl(\ell=1,2,3,\cdots ,{\alpha+\beta\over 2}\Bigr)$ \\
(b) double poles at $t=-{\alpha+\beta+1\over 2}-\ell\quad(\ell=0,1,2,3,\cdots,)$\\
(c) double poles at $t=-{\alpha+\beta+1+\ell\over 2}\quad(\ell=1,3,5,\cdots,)$\\
As it may be shown readily that the contributions from the double poles given in (b) and (c) vanish as $v\to 0$. Therefore, the contribution from only the simple poles give the result for $I_1$ will be
\begin{eqnarray}
I_1&=&{\sqrt{\pi}\over 2}\sum_{\ell=1}^{{1\over 2}(\alpha+\beta)}{\Gamma(\ell)\Gamma(\alpha+\beta-2\ell+1)\Gamma({1\over2}(\alpha+\beta)-\ell+{1\over 2})\over \Gamma(\alpha+\beta-\ell+1)\Gamma(\alpha-\ell+1)\Gamma(\beta-\ell+1)}\nonumber\\
& &\times{J_{-{\alpha+\beta\over 2}+\ell}(\kappa v)\over (\kappa v/2)^{{1\over2}(\alpha+\beta)-\ell}\Bigl[\pi(-1)^{{1\over 2}(\alpha+\beta)-\ell}\Bigr]}\Bigl({\kappa\over 2}\Bigr)^{\alpha+\beta-2\ell}
\end{eqnarray}
Considering that ${\alpha+\beta\over 2}-\ell$ is integer, we have
\begin{eqnarray}
\lim_{v\to 0}\Bigl({\kappa v\over 2}\Bigr)^{-{1\over 2}(\alpha+\beta)+\ell}J_{-{1\over 2}(\alpha+\beta)+\ell}(\kappa v)={(-1)^{{1\over2}(\alpha+\beta)-\ell}\over \Gamma({1\over 2}(\alpha+\beta)-\ell+1)}
\end{eqnarray}
Changing the index of summation from $\ell$ to $p={\alpha+\beta\over 2}-\ell$, we finally obtain the expression of the series $I_1$
\begin{eqnarray}
I_1&=&{1\over 2\sqrt{\pi}}\sum_{p=0}^{(\alpha+\beta-1)/2}{\Gamma({1\over 2}(\alpha+\beta)-p)\Gamma(2p+1)\Gamma(p+{1\over 2})\over \Gamma({1\over 2}(\alpha+\beta)+p+1)\Gamma({1\over 2}(\alpha-\beta)+p+1)}\nonumber\\
& &\times{(\kappa/2)^{2p}\over\Gamma({1\over 2}(\beta-\alpha)+p+1)\Gamma(p+1)}
\end{eqnarray}\\
\smallskip
\textbf{(b) Evaluation of the Integral $I_2$}\\
\smallskip
The integrand of the integral $I_2$ has simple poles at $t=p(p=0,1,2,,\cdots,)$ and the result becomes the same form as the previous case.
\begin{eqnarray}
I_2&=&-j{\sqrt{\pi}\over 2}\sum_{p=0}^\infty{\Gamma(\alpha+\beta+2p+1)\Gamma[{1\over2}(\alpha+\beta)+p+{1\over 2}]\over \Gamma(p+1)\Gamma(\alpha+\beta+p+1)\Gamma(\alpha+\beta+1)}\nonumber\\
& &\times{1\over \Gamma(\beta++p+1)\Gamma[{1\over2}(\alpha+\beta)+p+{1\over2}]}\Bigl({k\over2}\Bigr)^{\alpha+\beta+2p}
\end{eqnarray}
\smallskip
\textbf{(c) Evaluation of the Integral $I_3$}\\

\smallskip
This integral has simple at $t=p+{1\over2}(p=0,1,2,\cdots,)$ and the result is given by
\begin{eqnarray}
I_3&=&-{\sqrt{\pi}\over 2}\sum_{p=0}^\infty{(-1)^p\Gamma(\alpha+\beta+2p+2)\Gamma[{1\over2}(\alpha+\beta)+p+1]\over\Gamma(\alpha+\beta+p+{3\over 2})\Gamma(\alpha+p+{3\over2})\Gamma(\beta+p+{3\over2})}\nonumber\\
& &\times{1\over\Gamma(p+{3\over2})\Gamma[{1\over2}(\alpha+\beta+1)+p+1]}\Bigl({\kappa\over2}\Bigr)^{\alpha+\beta+2p+1}
\end{eqnarray}

\section{Series Solution of the Integral $G_2(\alpha,\beta;\kappa)$}

Considering an evaluation of the integral defined by
\begin{eqnarray}
G_2(\alpha,\beta;\kappa)&=&\int_0^\infty{J_\alpha(\xi) J_\beta(\xi)\over \xi^2\sqrt{\xi^2-\kappa^2}}d\xi
\end{eqnarray}
we defined Eq.(3.26) as the limiting value of the integral, given by
\begin{eqnarray}
G_2(\alpha,\beta;\kappa)&=&\lim_{v\to 0}\int_0^\infty {J_\alpha(\xi)J_\beta(\xi)\over\xi^2\sqrt{\xi^2-\kappa^2}}\exp[-\sqrt{\xi^2-\kappa^2}v]d\xi
\end{eqnarray}
Using Eq.(3.10) and (3.11), we have the above equation can be written as
\begin{eqnarray}
G_2(\alpha,\beta;\kappa)&=&{1\over\pi^2 j}\int_{-j\infty-\xi}^{+j\infty-\xi}{\Gamma(-t)dt\over\Gamma(\alpha+\beta+t+1)}\int_0^{{\pi\over2}}\cos^{\alpha+\beta+2t}\theta\cos[(\alpha-\beta)\theta]d\theta\nonumber\\
& &\times\int_0^\infty{1\over\sqrt{\xi^2-\kappa^2}}\xi^{\alpha+\beta+2t-2}\exp[-\sqrt{\xi^2-\kappa^2}v]d\xi
\end{eqnarray}
the integral with respect to $\theta$ and $\xi$ may be carried out i.e Eq.(3.13) and (3.14), the above equation becomes
\begin{eqnarray}
G_2(\alpha,\beta;\kappa)&=&{1\over16\sqrt{\pi}j}\int_{-j\infty-\xi}^{+j\infty-\xi}{\Gamma(-t)\over\Gamma(\alpha+\beta+t+1)}\nonumber\\
& &\times{\Gamma(\alpha+\beta+2t+1)\Gamma[{1\over2}(\alpha+\beta)+t-{1\over2}]\over\Gamma(\alpha+t+1)\Gamma(\beta+t+1)}{({\kappa/2})^{\alpha+\beta+2t-2}\over(\kappa v/2)^{(\alpha+\beta)/2+t-1}}\nonumber\\
& &\times\Bigl[-Y_{{1\over2}(\alpha+\beta)+t-1}(\kappa v)-jJ_{{1\over2}(\alpha+\beta)+t-1}(\kappa v)\Bigr]dt
\end{eqnarray}
The Neumann function in the above equation is replaced by using the relation, given in Eq.(3.16), $G_2(\alpha,\beta;\kappa)$ can be splitted into three parts, given by
\begin{eqnarray}
I_1&=&{1\over16\sqrt{\pi}}\int_{-j\infty-\xi}^{+j\infty-\xi}Q(t)\csc\Bigl[\Bigl({\alpha+\beta\over2}+t-1\Bigr)\pi]J_{-[{1\over2}(\alpha+\beta)+t-1]}(\kappa v)dt\\
I_2&=&-{1\over\sqrt{\pi}}\int_{-j\infty-\xi}^{+j\infty-\xi}Q(t)J_{{1\over2}(\alpha+\beta)+t-1}(\kappa v)dt\\
I_3&=&-{1\over16\sqrt{\pi}}\int_{-j\infty-\xi}^{+j\infty-\xi}Q(t)\cot\Bigl[\Bigl({\alpha+\beta\over2}+t-1\Bigr)\pi\Bigr]J_{{1\over2}(\alpha+\beta)+t-1}(\kappa v)dt\\
Q(t)&=&{\Gamma(-t)\Gamma(\alpha+\beta+2t+1)\Gamma[{1\over2}(\alpha+\beta)+t-{1\over2}]\over\Gamma(\alpha+\beta+t+1)\Gamma(\alpha+t+1)\Gamma(\beta+t+1)}\Bigl({\kappa\over2v})^{{1\over2}(\alpha+\beta)+t-1}
\end{eqnarray}
In the limit $|t|\to \infty$, the integral of $I_1$ approaches to $\exp[2t\ln(2/v)$, while those of $I_2$ and $I_3$ approach $\exp[-t\in v]$. Therefore, we may close the contour of $I_1$ in the left half plane for $v < 2$, and those of $I_2$ and $I_3$ in the right half plane. Since the evaluation of the integral depends on the indices $\alpha$ and $\beta$, we consider the following cases.\\
\smallskip
\textbf{(a) Evaluation of the Integral $I_1$}\\
\smallskip
The integrand of the integral $I_1$ has the singularities\\
(a) simple poles at $t=-{1\over2}-p\quad\Bigl(p=0,1,2,3,\cdots,{\alpha+\beta-1\over2})$\\
(b) double poles at $t=-{\alpha+\beta+1\over 2}(p=0,1,2,\cdots,)$\\
(c) double poles at $t=-{\alpha+\beta+1+p\over2}(p=1,3,5,\cdots,)$\\
As in the case (a), the contributions from the double poles vanish as $v\to 0$. The result is
\begin{eqnarray}
I_1&=&{1\over8\sqrt{\pi}}\sum_{p=0}^{(\alpha+\beta-3)/2}{\Gamma[{1\over2}(\alpha+\beta)-p-1]\Gamma(2p+3)\Gamma(p+{1\over2})\over\Gamma[{1\over2}(\alpha+\beta)+p+2]\Gamma[{1\over2}(\alpha-\beta)+p+2]}\nonumber\\
& &\times{1\over\Gamma[{1\over2}(\beta-\alpha)+p+2]\Gamma(p+1)}\Bigl({\kappa\over2}\Bigr)^{2p}
\end{eqnarray}\\
\smallskip
\textbf{(b) Evaluation of the Integral $I_2$}\\
\smallskip
The integrand of the integral $I_2$ has simple poles at $t=p(p=0,1,2,\cdots,)$ and the result becomes the same form as the previous one, that is
\begin{eqnarray}
I_2&=&-j{\sqrt{\pi}\over8}\sum_{p=0}^\infty{(-1)^p\Gamma(\alpha+\beta+2p+1)\Gamma[{1\over2}(\alpha+\beta)+p-{1\over2}]\over\Gamma(p+1)\Gamma(\alpha+\beta+p+1)\Gamma(\alpha+p+1)}\nonumber\\
& &\times{1\over\Gamma(\beta+p+1)\Gamma[{1\over2}(\alpha+\beta)+p]}\Bigl({\kappa\over2}\Bigr)^{\alpha+\beta+2p-2}
\end{eqnarray}\\
\smallskip
\textbf{(c) Evaluation of the Integral $I_3$}\\
\smallskip
The integrand of this integral has simple pole at $t=p+{1\over2}(p=0,1,2,\cdots,)$ and the result is given by
\begin{eqnarray}
I_3&=&-{\sqrt{\pi}\over8}\sum_{p=0}^\infty{(-1)^p\Gamma(\alpha+\beta+2p+2)\Gamma[{1\over2}(\alpha+\beta)+p]\over\Gamma(\alpha+\beta+p+{3\over2})\Gamma(\alpha+p+{3\over2})\Gamma(\beta+p+{3\over2})}\nonumber\\
& &\times{1\over\Gamma(p+{3\over2})\Gamma[{1\over2}(\alpha+\beta)+p+{1\over2}]}\Bigl({\kappa\over2}\Bigr)^{\alpha+\beta+2p-1}
\end{eqnarray}

\section{Series Solution of the Integral $K(\alpha,\beta;\kappa)$}

$K(\alpha,\beta;\kappa)$ is defined by 
\begin{eqnarray}
K(\alpha,\beta;\kappa)&=&\lim_{v\to 0}\int_0^\infty\sqrt{\xi^2-\kappa^2}{J_\alpha(\xi)J_\beta(\xi)\over\xi^2}\exp[-\sqrt{\xi^2-\kappa^2}v]d\xi
\end{eqnarray}
This integral may also be carried out as in $G(\alpha,\beta;\kappa)$. Here transforming $K(\alpha,\beta;\kappa)$ into
\begin{eqnarray}
K(\alpha,\beta;\kappa)&=&\lim_{v\to 0}{1\pi j}\int_{-j\infty-\xi}^{+j\infty-\xi}{\Gamma(-t)\Gamma(\alpha+\beta+2t+1)\over\Gamma(\alpha+\beta+t+1)\Gamma(\alpha+t+1)\Gamma(\beta+t+1)2^{\alpha+\beta+2t+1}}dt\nonumber\\
& &\times\int_0^\infty\sqrt{\xi^2-\kappa^2}\xi^{\alpha+\beta+2t-2}\exp(-\sqrt{\xi^2-\kappa^2}v)d\xi
\end{eqnarray}
Using the relation
\begin{eqnarray}
I&=&\int_0^\infty\sqrt{\xi^2-\kappa^2}\Bigl({\xi\over2}\Bigr)^{\alpha+\beta+2t-2}\exp(-\sqrt{\xi^2-\kappa^2}v)d\xi\nonumber\\
 &=&4\int_0^\infty{1\over\sqrt{\xi^2-\kappa^2}}\left\{\Bigl({xi\over2}\Bigr)^{\alpha+\beta+2t}-\Bigl({\kappa\over2}\Bigr)^2\Bigl({\xi\over2}\Bigr)^{\alpha+\beta+2t+2}\right\}\exp(-\sqrt{\xi^2-\kappa^2}v)d\xi\nonumber\\
&=&2\sqrt{\pi}\Bigl({\kappa\over2}\Bigr)^{\alpha+\beta+2t}\left[{\Gamma[{1\over2}(\alpha+\beta)+t+{1\over2}]\over(\kappa v/2)^{{1\over2}(\alpha+\beta)+t}}\left\{-Y_{{1\over2}(\alpha+\beta)+t}(\kappa v)-jJ_{{1\over2}(\alpha+\beta)+t}(\kappa v)\right\}\right.\nonumber\\
& &\left.-{\Gamma[{1\over2}(\alpha+\beta)+t-{1\over2}]\over(\kappa v/2)^{{1\over2}(\alpha+\beta)+t-1}}\left\{-Y_{{1\over2}(\alpha+\beta)+t-1}(\kappa v)-jJ_{{1\over2}(\alpha+\beta)+t-1}(\kappa v)\right\}\right]
\end{eqnarray}
$K(\alpha,\beta;\kappa)$ may be split into three parts
\begin{eqnarray}
K(\alpha,\beta;\kappa)&=&I_1+I_2+I_3\nonumber\\
I_1&=&{1\over4\sqrt{pi}j}\int_{-j\infty-\xi}^{+j\infty-\xi}P(t)\{Q_A(t)-Q_A(t-1)\}dt\nonumber\\
I_2&=&-{1\over4\sqrt{\pi}}\int_{-j\infty-\xi}^{+j\infty-\xi}P(t)\{Q_B(t)-Q_B(t-1)\}dt\nonumber\\
I_3&=&-{1\over4\sqrt{\pi}j}\int_{-j\infty-\xi}^{+j\infty-\xi}P(t)\{Q_C(t)-Q_C(t-1)\}dt\nonumber\\
\end{eqnarray}
where
\begin{eqnarray}
P(t)&=&{\Gamma(-t)\Gamma(\alpha+\beta+2t+1)\over\Gamma(\alpha+\beta+t+1)\Gamma(\alpha+t+1)\Gamma(\beta+t+1)}\Bigl({\kappa\over2}\Bigr)^{\alpha+\beta+2t}\\
Q_A(t)&=&\Gamma\Bigl({\alpha+\beta\over2}+t+{1\over2}\Bigr)\Big({\kappa v\over2}\Bigr)^{-{1\over2}(\alpha+\beta+2t)}\csc\Bigl({\alpha+\beta+2t\over2}\pi\Bigr)J_{-{1\over2}(\alpha+\beta+2t)}(\kappa v)\\
Q_B(t)&=&\Gamma\Bigl({\alpha+\beta\over2}+t+{1\over2}\Bigr)\Bigl({\kappa v\over2}\Bigr)^{-{1\over2}(\alpha+\beta+2t)}J_{-{1\over2}(\alpha+\beta+2t)}(\kappa v)\\
Q_C(t)&=&\Gamma\Bigl({\alpha+\beta\over2}+t+{1\over2}\Bigr)\Bigl({\kappa v\over2}\Bigr)^{-{1\over2}(\alpha+\beta+2t)}\cot\Bigl({1\over2}(\alpha+\beta+2t)\pi\Bigr)\nonumber\\
& &\times J_{{1\over2}(\alpha+\beta+2t)}(\kappa v)
\end{eqnarray}
The method of solving the above integrals are the same as discussed in the previous sections. Here we have show the result, that is
\begin{eqnarray}
I_1&=&-{1\over 4\sqrt{\pi}}\sum_{p=0}^{{1\over2}(\alpha+\beta)-1}{\Gamma[{1\over2}(\alpha+\beta)]\Gamma(2p+1)\Gamma(p-{1\over2})\over\Gamma(p+1)\Gamma[{1\over2}(\alpha+\beta)+p+1]\Gamma[{1\over2}(\alpha+\beta)+p+1]}\nonumber\\
& &\times{1\over\Gamma[{1\over2}(\alpha+\beta)+p+1]}\Bigl({\kappa\over2}\Bigr)^{2p}\\
I_2&=&j{\sqrt{\pi}\over 4}\sum_{p=0}^\infty{(-1)^p\Gamma(\alpha+\beta+2p+1)\Gamma[{1\over2}(\alpha+\beta)+p-{1\over2}]\over\Gamma(p+1)\Gamma(\alpha+\beta+p+1)\Gamma[{1\over2}(\alpha+\beta)+p+1]}\nonumber\\
& &\times{1\over\Gamma(\alpha+p+1)\Gamma(\beta+p+1)}\Bigl({\kappa\over2}\Bigr)^{\alpha+\beta+2p}\\
I_3&=&{\sqrt{\pi}\over4}\sum_{p=0}^\infty{(-1)^p\Gamma(\alpha+\beta+2p+2)\Gamma[{1\over2}(\alpha+\beta)+p]\over\Gamma(\alpha+\beta+p+{3\over2})\Gamma(p+{3\over2})\Gamma[{1\over2}(\alpha+\beta)+p+{3\over2}]}\nonumber\\
& &\times{1\over\Gamma(\alpha+p+{3\over2})\Gamma(\beta+p+{3\over2})}\Bigl({\kappa\over2}\Bigr)^{\alpha+\beta+2p+1}
\end{eqnarray}
It is noted that $K(\alpha,\beta;\kappa)$ is related to $G(\alpha,\beta;\kappa)$ and $G_2(\alpha,\beta;\kappa)$ by the relation given by
\begin{eqnarray}
K(\alpha,\beta;\kappa)&=&G(\alpha,\beta;\kappa)-G_2(\alpha,\beta;\kappa)
\end{eqnarray}
\section{Numerical Integration}
\paragraph{} Numerical integration is the approximate computation of an integral using numerical techniques. The numerical computation of an integral is sometime called quadrature.The most straightforward numerical integration technique uses the Newton-Cotes formulas, which approximate a function tabulated at a sequence of regularly spaced intervals by various degree polynomials. If the endpoints are tabulated, then the 2- and 3-point formulas are called the trapezoidal rule and the Simpson's rule, respectively. The 5-point formula is called Boole's rule. A generalization of the trapezoidal rule is Romberg integration, which can yield accurate results for many fewer function evaluations.\\
\indent If the function are know analytically instead of being tabulated at equally spaced intervals, the best numerical method of integration is call Gaussian quadrature. By picking the abscissas at which to evaluate the function, Gaussian quadrature produces the post accurate approximation possible. However, given the speed of modern computers, the additional complication of Gaussian quadrature formalism after makes it less desirable that simply brute-force calculating twice as many points on a regular grid. Here we have computed the integrals by using the Gauss-Legendre quadrature, which is simply a check for the series solution of the integrals.
\subsection{Numerical Integration of $G(\alpha,\beta;\kappa)$}
\paragraph{}
The integral $G(\alpha,\beta;\kappa)$ can be also computed numerically by using the Gauss-Legendre quadrature. Here we set
\begin{eqnarray}
G(\alpha,\beta;\kappa)&=&G_{re}(\alpha,\beta;\kappa)-jG_{im}(\alpha,\beta;\kappa)
\end{eqnarray}
The imaginary part of $G(\alpha,\beta;\kappa)$ is computed from
\begin{eqnarray}
G_{im}(\alpha,\beta;\kappa)&=&\int_0^\kappa{1\over\sqrt{\kappa^2-\xi^2}}J_\alpha(\xi)J_\beta(\xi)d\xi\nonumber\\
&=&2\int_0^{\sqrt{\kappa}}{1\over\sqrt{2\kappa-\eta^2}}J_\alpha(\kappa-\eta^2)J_\beta(\kappa-\eta^2)d\eta
\end{eqnarray}
The real part is split into three parts as
\begin{eqnarray}
G_{re}(\alpha,\beta;\kappa)&=&\int_\kappa^\infty{1\sqrt{\kappa^2-\xi^2}}J_\alpha(\xi)J_\beta(\xi)d\xi\equiv G_1+G_2+G_3
\end{eqnarray}
where
\begin{eqnarray}
G_1&=&\int_\kappa^{3\kappa}{1\over\sqrt{\kappa^2-\xi^2}}J_\alpha(\xi)J_\beta(\xi)d\xi=2\int_0^{\sqrt{2\kappa}}{1\over\sqrt{2\kappa+\eta^2}}J_\alpha(\kappa+\eta^2)J_\beta(\kappa+\eta^2)d\eta\\
G_2&=&\int_{3\kappa}^X{1\over\sqrt{\kappa^2-\xi^2}}J_\alpha(\xi)J_\beta(\xi)d\xi\\
G_3&=&\int_0^\infty{1\over\sqrt{\kappa^2-\xi^2}}J_\alpha(\xi)J_\beta(\xi)d\xi=\int_X^\infty{1\over\xi}\Bigl(1+{\kappa^2\over2\xi^2}\Bigr)J_\alpha(\xi)J_\beta(\xi)d\xi\nonumber\\
&\simeq&{1\over\pi}\Big\{{1\over X}+{1\over3X^3}\Bigl[{\kappa^2\over2}-\Bigl(a_2+b_2-a_1b_1\Bigr)\Bigr]\Bigr\}\cos{(\alpha-\beta)\pi\over2}+{1\over\pi}{a_1-b_1\over2X^2}\sin{(\alpha-\beta)\pi\over2}\nonumber\\
&-&{1\over\pi}{1\over2X^2}\sin\Bigl(2X-{\alpha+\beta+1\over2}\pi\Bigr)-{a_1+b_1\over\pi}{1\over2X^3}\cos\Bigl(2X-{\alpha+\beta+1\over2}X\Bigr)
\end{eqnarray}
where
\begin{eqnarray}
a_1={4\alpha^2-1\over8},\qquad a_2={(4\alpha^2-1)(4\alpha^2-9)\over128}\nonumber\\
b_1={4\beta^2-1\over8},\qquad  b_2={(4\beta^2-1)(4\beta^2-9)\over128}
\end{eqnarray}
\subsection{Numerical Integration of the Integral $G_2(\alpha,\beta;\kappa)$}
\paragraph{}
We set again
\begin{eqnarray}
G_2(\alpha,\beta;\kappa)&=&G_{2re}(\alpha,\beta;\kappa)-jG_{2im}(\alpha,\beta;\kappa)
\end{eqnarray}
The imaginary part of $G$ is computed from
\begin{eqnarray}
G_{2im}(\alpha,\beta;\kappa)&=&\int_0^\kappa{1\over\sqrt{\kappa^2-\xi^2}}J_\alpha(\xi)J_\beta(xi)d\xi\nonumber\\
&=&2\int_0^{\sqrt{\kappa}}{1\over(\kappa-\eta^2)^2\sqrt{2\kappa-\eta^2}}J_\alpha(\kappa-\eta^2)J_\beta(\kappa-\eta^2)d\eta
\end{eqnarray}
The real part is split into three parts as
\begin{eqnarray}
G_{2re}(\alpha,\beta;\kappa)&=&\int_\kappa^\infty{1\over\xi^2\sqrt{\kappa^2-\xi^2}}J_\alpha(\xi)J_\beta(\xi)d\xi\equiv G_{21}+G_{22}+G_{23}\end{eqnarray}
where
\begin{eqnarray}
G_{21}&=&\int_\kappa^{3\kappa}{1\over\xi^2\sqrt{\kappa^2-\xi^2}}J_\alpha(\xi)J_\beta(\xi)d\xi\nonumber\\
&=&2\int_0^{\sqrt{2\kappa}}{1\over(\kappa+\eta^2)^2\sqrt{2\kappa+\eta^2}}J_\alpha(\kappa+\eta^2)J_\beta(\kappa+\eta^2)d\eta\\
G_{22}&=&\int_{3\kappa}^X{1\over\xi^2{\kappa^2-\xi^2}}J_\alpha(\xi)J_\beta(\xi)d\xi\\
G_{23}&=&\int_\kappa^\infty{1\over\xi^2\sqrt{\kappa^2-\xi^2}}J_\alpha(\xi)J_\beta(\xi)d\xi=\int_X^\infty{1\over\xi^3}\Bigl(1+{\kappa^2\over2\xi^2}\Bigr)J_\alpha(\xi)J_\beta(\xi)d\xi\nonumber\\
&\simeq&{1\over\pi}\Bigl\{{1\over3X^3}+{1\over5X^5}\Bigl[{\kappa^2\over2}-\Bigl(a_2+b_2-a_1b_1\Bigr)\Bigr]\Bigr\}\cos{(\alpha-\beta)\pi\over2}+{1\over\pi}{a_1-b_1\over4X^4}\sin{(\alpha-\beta)\pi\over2}\nonumber\\
&-&{1\over\pi}{1\over4X^4}\sin\Bigl(2X-{(\alpha+\beta+1)\pi\over2}\Bigr)-{1\over\pi}{a_1+b_1\over2X^5}\cos\Bigl(2X-{\alpha+\beta+1\over2}\pi\Bigr)
\end{eqnarray}
where $a_1,b_1$ and $a_2,b_2$ are defined in Eq.(3.55).
\subsection{Numerical Integration of the Integral $K(\alpha,\beta;\kappa)$}
\paragraph{}We set
\begin{eqnarray}
K(\alpha,\beta;\kappa)&=&K_{re}(\alpha,\beta;\kappa)+jK_{im}(\alpha,\beta;\kappa)
\end{eqnarray}
The imaginary part of $K(\alpha,\beta;\kappa)$ is computed from
\begin{eqnarray}
K_{im}(\alpha,\beta;\kappa)&=&\int_o^\infty{\sqrt{\kappa^2-\xi^2}\over\xi^2}J_\alpha9\xi)J_\beta(\xi)d\xi\nonumber\\
&=&2\int_0^{\sqrt{\kappa}}{\sqrt{2\kappa-\eta^2}\over(\kappa-\eta^2)^2}\eta^2 J_\alpha(\kappa-\eta^2)J_\beta(\kappa-\eta^2)d\eta
\end{eqnarray}
The real part is split into three parts as
\begin{eqnarray}
K_{re}(\alpha,\beta;\kappa)&=&\int_\kappa^\infty{\sqrt{\kappa^2-\xi^2}\over\xi^2}J_\alpha(\xi)J_\beta(\alpha)d\xi\equiv K_1+K_2+K_3
\end{eqnarray}
where
\begin{eqnarray}
K_1&=&\int_\kappa^{3\kappa}{\sqrt{\xi^2-\kappa^2}\over\xi^2}J_\alpha(\xi)J_\beta(\xi)d\xi\nonumber\\
&=&2\int_0^{\sqrt{2\kappa}}{\sqrt{2\kappa+\eta^2}\over(\kappa+\eta^2)}\eta^2 J_\alpha(\kappa+\eta^2)J_\beta(\kappa+\eta^2)d\eta\\
K_2&=&\int_{3\kappa}^X{\sqrt{\xi^2-\kappa^2}\over\xi^2}J_\alpha(\xi)J_\beta(\xi)d\xi\\
K_3&=&\int_{X}^\infty{\sqrt{\xi^2-\kappa^2}\over\xi^2}J_\alpha(\xi)J_\beta(\xi)d\xi\nonumber\\
&=&\int_X^\infty{1\over\xi}\Bigl(1-{\kappa^2\over2\xi^2}\Bigr)J_\alpha(\xi)J_\beta(\xi)d\xi\nonumber\\
&\simeq&{1\over\pi}\Bigl\{{1\over X}-{1\over3X^3}\Bigl[{\kappa^2\over2}+\Bigl(a_2+b_2-a_1b_1\Bigr)\Bigr]\Bigr\}\cos{(\alpha-\beta)\pi\over2}+{1\over\pi}{a_1-b_1\over2X^2}\sin{(\alpha-\beta)\pi\over2}\nonumber\\
&-&{1\over\pi}{1\over2X^2}\sin\Bigl(2X-{\alpha+\beta+1\over2}\pi\Bigr)-{1\over\pi}{a_1+b_1\over2X^3}\cos\Bigl(2X-{\alpha+\beta+1\over2}\pi\Bigr)
\end{eqnarray}
\indent To verify the validity of the series expression of $G(\alpha,\beta;\kappa)$, $G_2(\alpha,\beta;\kappa)$ and $K(\alpha,\beta;\kappa)$ which are derived in the previous section, we perform numerical computation and the result are compared with those obtained by direct numerical integration. The agreement is fairly good.

\chapter{Numerical Discussion}

\paragraph{} With the formulation developed in the previous section, we have preformed the computation of the diffracted field for a disk with $\kappa a=5.0$, $\kappa a=9.0$, $\kappa a =12.0$ and $\kappa a=16.0$. For all cases computed, the thickness of the perfectly conducting disk is considered as infinitesimal. In order to check the effectiveness of the present method based on the Kobayashi potential, the numerical data for all the  cases mention above are compared with the results obtained using the physical optics (PO) method.\\

\section{Computations of Matrix Elements and Radiation Pattern}
\paragraph{} A first step in obtaining numerical results of the physical quantities is to compute the matrix elements defined in Eq.(2. ) and (2. ).These are infinite integral and can be computed by using the series solution of the infinite integral. From this, we obtained the expansion coefficient by using the matrix inversion. Once the numerical result for the expansion coefficients are obtained, the radiation patterns are computed from Eq(2.80) and (2.84). The numerical results of the far-filed pattern diffracted by a perfectly conducting plate for normal incidence are shown in figures below. The ordinate $\sigma$ denotes the power pattern or the differential scattering cross section. To verify the validity of the present computation, we show the result produces by the physical optics (PO) in these figures.

\section{Conclusion}
\paragraph{} We have formulated the electromagnetic field diffracted by circular conducting disk and circular hole in the conducting plate, when plane wave impinges on the obstacle, by using the method of the Kobayashi potential. We have write the expression for the vector potential by using the Helmholtz equation, in the form of the spectrum function. From these vector potential the corresponding electromagnetic field are derived. These expression are then written in the matrix form for the sack of simplicity. Finally the field expressions are expanded in the form of double series and each summand satisfies desired edge conditions as well as a part of other required boundary conditions.The mathematical formulation involves dual integral equations derived from the potential integral and the boundary condition on the plane where the disk or hole is located. The dual integral has been solved by using the Weber-Schafheitlin's discontinuous integral and the the orthogonal property of the Jacobi's polynomials. From this we obtained the weighting function in form of the matrix equations. Expansion coefficients are determined from the solution of the matrix equation and the simple series expansions for the matrix elements are derived. The matrix equation involves the infinite integral, which has been transform into infinite series expansion. These series expansion are more  convenient for numerical computation.\\
\indent We derive the far field expressions by two different ways. One method is to evaluate the field radiated from the current density induced on the disk and the second method is to evaluate the expression of the vector potential directly by applying the stationary phase method of integration.We presented the numerical results of the far diffracted field pattern by normal incidence, which is fairly agree with physical optics method (PO). The main problem in computational work is the numerical solution of the infinite integral which involve in the matrix equation for the expansion coefficient. Here we present the infinite series solution of the these infinite integral. Furthermore the the direct numerical solution of these integral are also presented for the compression with the series solution. The present method promises applicability to a wide class of problem such as, circular hole in the thick conducting plate, flanged circular resonator, and so on.

\pagebreak

\vfill
\bigskip
\begin{figure}[h]
\begin{center}
\includegraphics[width=7in, height=5.25in]{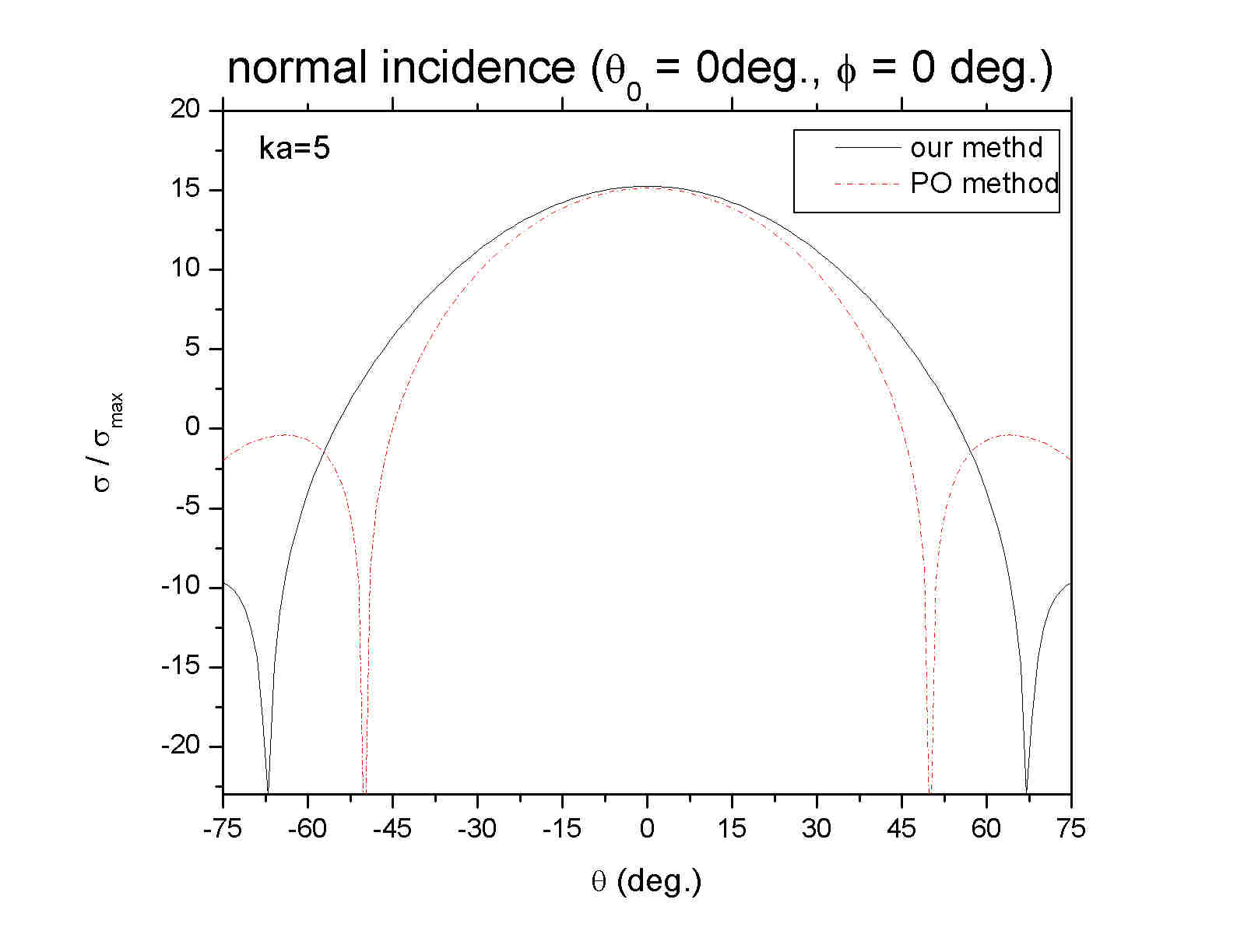}
\caption{Far-field pattern diffracted by a perfectly conducting plate for normal incidence $(\theta_0=0^0, \phi=0^0)$. Disk is $\kappa =\kappa a=5$.\hfill}
\end{center}
\end{figure}
\vfill

\vfill
\bigskip
\begin{figure}[h]
\begin{center}
\includegraphics[width=7in, height=5.25in]{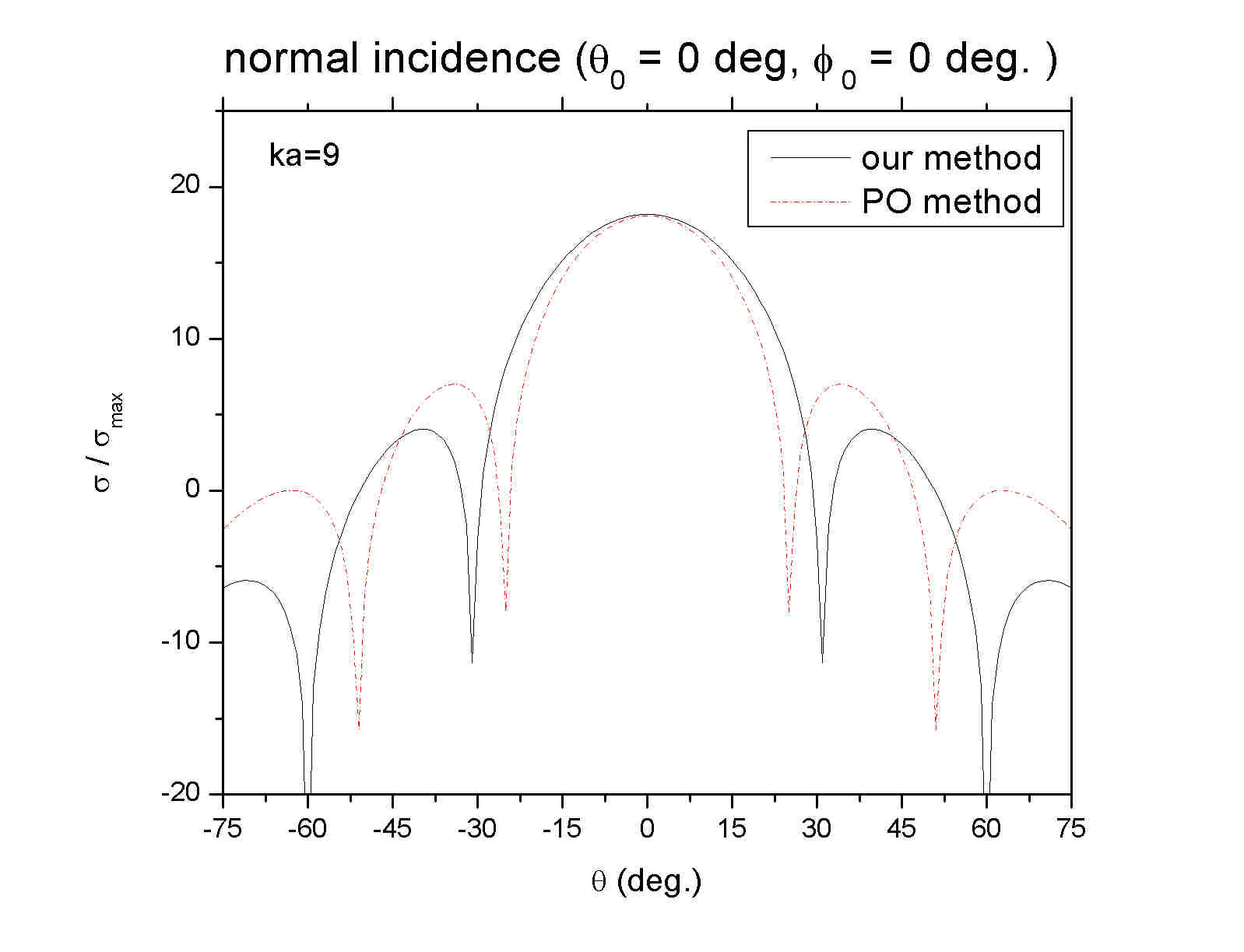}
\caption{Far-field pattern diffracted by a perfectly conducting plate for normal incidence $(\theta_0=0^0, \phi=0^0)$. Disk is $\kappa =\kappa a=9$.\hfill}
\end{center}
\end{figure}
\vfill

\vfill
\bigskip
\begin{figure}[h]
\begin{center}
\includegraphics[width=7in, height=5.25in]{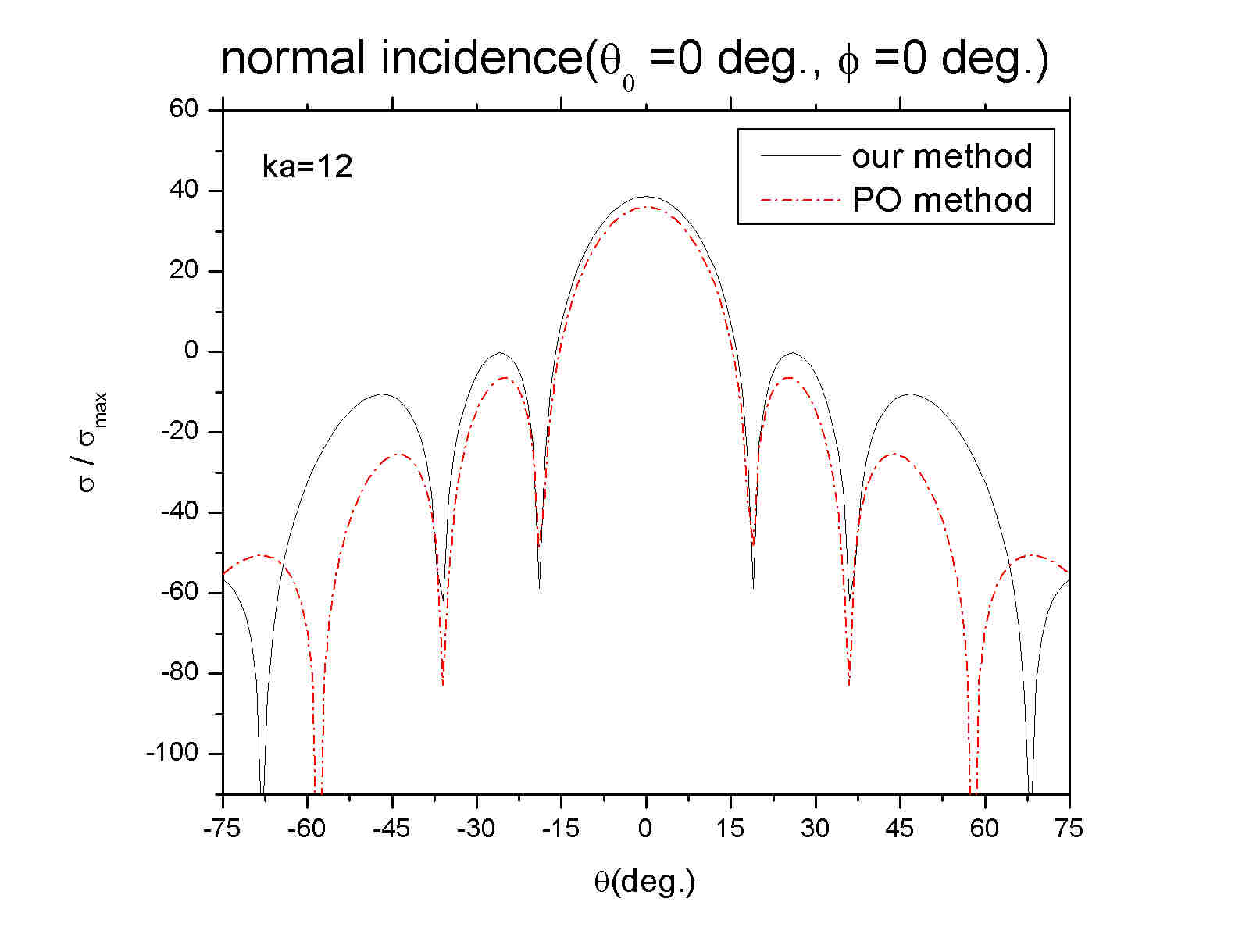}
\caption{Far-field pattern diffracted by a perfectly conducting plate for normal incidence $(\theta_0=0^0, \phi=0^0)$. Disk is $\kappa =\kappa a=12$.\hfill}
\end{center}
\end{figure}
\vfill

\vfill
\bigskip
\begin{figure}[h]
\begin{center}
\includegraphics[width=7in, height=5.25in]{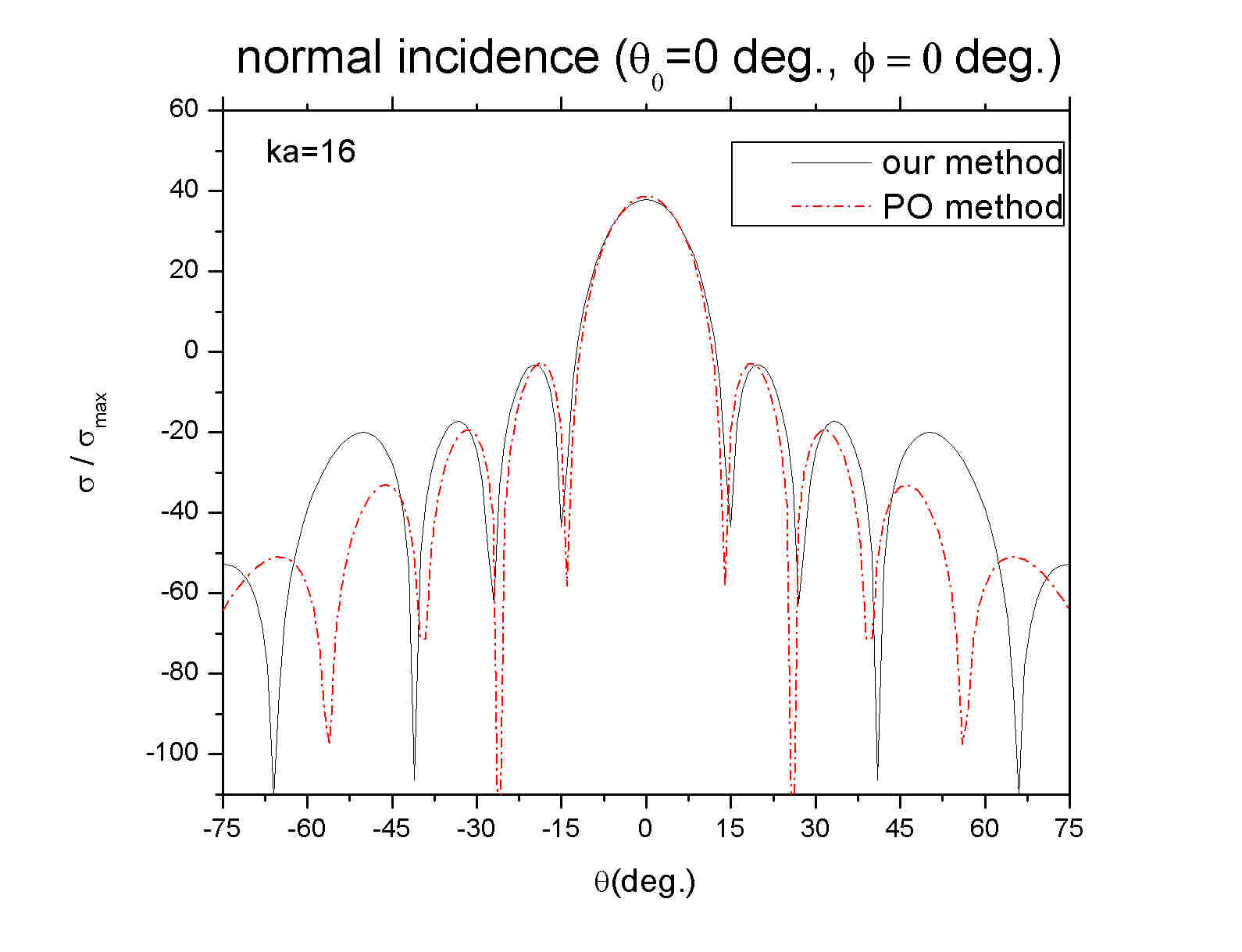}
\caption{Far-field pattern diffracted by a perfectly conducting plate for normal incidence $(\theta_0=0^0, \phi=0^0)$. Disk is $\kappa =\kappa a=16$.\hfill}
\end{center}
\end{figure}
\vfill


\appendix
\chapter{ Vector Hankel Transform}
\paragraph{}
The vector Hankel transform is the generalization of the conventional Hankel transform and it transform a vector function from one space to a vector function  in another space. This was introduced by Chew and Kong [41],[42] to analyse the disk resonator and antenna. Here we will reproduce the proof by assuming the transform pair
$$\left[\matrix{f_1(\rho_a)\cr f_1(\rho_a)}\right]=\int_0^\infty\Bigl[H^\pm(\xi\rho_a)\Bigr]\left[\matrix{F_1(\xi)\cr F_2(\xi)}\right]\xi d\xi\eqno(A1)$$
$$\left[\matrix{ F_1(\xi)\cr F_2(\xi)}\right]=\int_0^\infty\Bigl[H^\pm(\xi\rho-a)\Bigr]\left[\matrix{f_1(\rho_a)\cr f_2(\rho_a)}\right]\rho_a d\rho_a\eqno(A2)$$
where
$$\Bigl[H^\pm(\xi\rho_a)\Bigr]=\left[\matrix{J_m^\prime(\xi\rho_a)\quad\pm{m\over\xi\rho_a}J_m(\xi\rho_a)\cr\pm{m\over\xi\rho_a}J_m(\xi\rho_a)\quad J_m^\prime(\xi\rho_a)}\right]\eqno(A3)$$
If Eq.(A1) is true, it implies
$$\left[\matrix{f_1(\rho_a)\cr f_2(\rho_a)}\right]=\int_0^\infty \xi d\xi\int_0^\infty\rho_a^\prime d\rho_a^\prime\Bigl[H^\pm(\xi\rho_a)\Bigr]\Bigl[H^\pm(\xi\rho_a^\prime)\Bigr]\left[\matrix{f_1(\rho_a^\prime)\cr f_2(\rho_a^\prime)}\right]\eqno(A4)$$
Eq.(A4) can be proved by interchanging the order of integration. The diagonal element of $\Bigl[H^\pm(\xi\rho_a)\Bigr]\Bigl[H^\pm(\xi\rho_a^\prime)\Bigr]$ become
$$\xi J_m^\prime(\xi\rho_a)J_m^\prime(\xi\rho_a^\prime)+{m^2\over \xi\rho_a\rho_a^\prime}J_m(\xi\rho_a)J_m(\xi\rho_a^\prime)=\xi J_{m+1}(\xi\rho_a)J_{m+1}(\xi\rho_a^\prime)+{m\over\rho_a\rho_a^\prime}{d\over d\xi}J_m(\xi\rho_a)J_m(\xi\rho_a^\prime)$$
Using the closure relation of the Hankel transform, given by
$$\int_0^\infty \xi J_{m+1}(\xi\rho_a)J_{m+1}(\xi\rho_a^\prime)d\xi={\delta(\rho_a-\rho_a^\prime)\over \rho_a^\prime}$$
the first and second term evaluates to zero. Similarly, the off diagonal elements can be written as
$${m\over\rho_a^\prime}J_m^\prime(\xi\rho_a)J_m(\xi\rho_a^\prime)+{m\over \rho_a}J_m(\xi\rho_a)J_m^\prime(\xi\rho_a^\prime)={m\over \rho_a\rho_a^\prime}J_m(\xi\rho_a)J_m(\xi\rho_a^\prime)$$
which integrate to give zero. Consequently Eq.(A4) becomes
\begin{eqnarray*}
\int_0^\infty \xi d\xi\int_0^\infty\rho_a^\prime d\rho_a^\prime\Bigl[H^\pm(\xi\rho_a)\Bigr]\Bigl[H^\pm(\xi\rho_a^\prime)\Bigr]\left[\matrix{f_1(\rho_a^\prime)\cr f_2(\rho_a^\prime)}\right]\\
=\int_0^\infty \rho_a^\prime d\rho_a^\prime\left[\matrix{{\delta(\rho_a-\rho_a^\prime)\over\rho_a^\prime}\quad 0\cr 0\qquad{\delta(\rho_a-\rho_a^\prime)\over \rho_a^\prime}}\right]\left[\matrix{f_1(\rho_a^\prime)\cr f_2(\rho_a^\prime)}\right]=\left[\matrix{f_1(\rho_a)\cr f_2(\rho_a)}\right] & &(A5)
\end{eqnarray*}
which complete the proof.

\appendix
\chapter{Weber-Schafheitlin's Discontinuous Integral}
\paragraph{}The Weber-Schafheitlin's discontinuous integral is defined by
\begin{eqnarray}
 W(\mu,\nu,\lambda;r)=\int_0^\infty{J_\mu(ru)J_\nu(u)\over u^\lambda}d \mu
\end{eqnarray}
To solve the above integral, we used the relations
\begin{eqnarray}
J_\nu(x)=\sum_{m=0}^\infty(-1)^m{({1\over 2}x)^{\nu+2m}\over m!\Gamma(m+\nu+1)}
\end{eqnarray}
and
\begin{eqnarray}
\int_0^\infty\exp[-at]J_\nu(bt)t^{\mu-1}dt\nonumber\\
&=&\sum_{m=0}^\infty(-1)^m{({1\over 2}b)^{\mu+2m}\over m!\Gamma(\nu+m+1)}\int_0^\infty t^{\mu+\nu+2m-1}\exp[-at]dt\nonumber\\
&=&\sum_{m=0}^\infty (-1)^m{({1\over 2}b)^{\nu+2m}\over m!\Gamma(\nu+m+1)}{\Gamma(\mu+\nu+2m)\over a^{\mu+\nu+2m}}
\end{eqnarray}
\pagebreak
Hence we can write
\begin{eqnarray}
W(\mu,\nu,\lambda;r)\nonumber\\
&=&{r^\mu\Gamma[{1\over2}(\mu+\nu-\lambda+1)]\over 2^\lambda\Gamma[{1\over2}(-\mu+\nu+\lambda+1)]\Gamma(\nu+1)}\nonumber\\
& &\times F\Bigl[{\mu+\nu-\lambda+1\over 2},{\mu-\nu-\lambda+1\over 2},\mu+1;r^2\Bigr]\quad 0\leq r<1\nonumber\\
&=&{\Gamma[{1\over 2}(\mu+\nu-\lambda+1)]\over 2^\lambda r^{\nu-\lambda+1}\Gamma[{1\over 2}(\mu-\nu+\lambda+1)]\Gamma(\nu+1)}\nonumber\\
& &\times F\Bigl[{\mu+\nu-\lambda+1\over 2},{-\mu+\nu-\lambda+1\over 2},\nu+1;{1\over r^2}\Bigr]\quad r\geq1
\end{eqnarray}
\indent If there exists a relation $\mu-\nu+\lambda=-2m-1(m=0,1,2,\cdots ,)$ among the parameters, the function becomes $W(\mu,\nu,\lambda;r)=0$ for $r>1$ because of the properties of the gamma function$|\Gamma(-n)|\to\infty$ (n: positive integer). For $r$ in the range $0\leq r<1$, $W(\mu,\nu,\lambda;r)$ takes finite value.

\appendix
\chapter{Expansion of the Bessel Function by Jacobi's Polynomials}
\section{Projection of $J_{m\pm1}(\xi\sqrt{x})$ into $\nu^m_n(x)$}
\paragraph{} 
We project $J_m^\prime(\xi\sqrt{x})$ and ${m\over \xi\sqrt{x}}J_m(\xi\sqrt{x})$ into functional space with element $\nu^m_n(x)$. To evaluate the components of each element , we need the following relation
\begin{eqnarray}
I_1&=&\int_0^1 x^{{1\over 2}(m+1)}(1-x)^{{1\over 2}}J_{m-1}(\xi\sqrt{x})\nu^m_p(x)dx\nonumber\\
&=&\sum_{q=0}^\infty{\sqrt{8}(2p+m+{1\over 2})\Gamma(q+m+{1\over 2})\over \Gamma(q+1)\Gamma(m)}{J_{2q+m+{1\over 2}}(\xi)\over \xi^{{3\over 2}}}I^{m-1,m}_{q,p}\nonumber\\
&=&{\sqrt{8}\Gamma(m+1)\Gamma(p+{3\over 2})\over (2p+m+{3\over 2})\Gamma(p+m)}{J_{2p+m+{1\over 2}}\over\xi^{{3\over 2}}}-{\sqrt{8}\Gamma(m+1)\Gamma(p+{5\over 2})\over(2p+m+{3\over 2})\Gamma(p+m+1)}{J_{2p+m+{5\over 2}}\over \xi^{{3\over 2}}}\nonumber\\
&=&{\sqrt{8}\Gamma(m+1)\Gamma(p+{3\over 2})\over (2p+m+{3\over 2})\Gamma(p+m+1)}\left[(p+m){J_{22n+m+{1\over 2}}(\xi)\over \xi^{{3\over 2}}}-\Bigl(p+{3\over 2}\Bigr){J_{2p+m+{5\over 2}}(\xi)\over \xi^{{3\over 2}}}\right]\\
I_2&=&\int_0^1 x^{{1\over 2}(m+1)}(1-x)^{{1\over 2}}J_{m+1}(\xi\sqrt{x})\nu^m_p(x)dx\nonumber\\
&=&\sum_{q=0}^\infty {\sqrt{8}(2p+m+{5\over 2})\Gamma(q+m+{5\over 2})\over \Gamma(q+1)\Gamma(m+2)}{J_{2q+m+{5\over 2}}(\xi)\over \xi^{{3\over 2}}}I^{m+1,m}_{q,p}\nonumber\\
&=&-{\sqrt{8}p\gamma(m+1)\Gamma(m+{3\over 2})\over(2p+m+{3\over 2})\Gamma(p+m+1)}{J_{2p+m+{1\over 2}}(\xi)\over \xi^{{3\over 2}}}\nonumber\\
& &+{\sqrt{8}(p+m+{3\over 2})\Gamma(m+1)\Gamma(m+1)\Gamma(p+{3\over 2})\over (2p+m+{3\over 2})\Gamma(p+m+1)}{J_{2p+m+{5\over 2}}(\xi)\over \xi^{{3\over 2}}}\nonumber\\
&=&{\sqrt{8}\Gamma(m+1)\Gamma(p+{3\over 2})\over(2p+m+{3\over 2})\Gamma(p+m+1)}\left[-p{J_{2p+m+{1\over 2}}(\xi)\over \xi^{{3\over 2}}}+\Bigl(p+m+{3\over 2}\Bigr){J_{2p+m+{5\over 2}}(\xi)\over \xi^{{3\over 2}}}\right]\\
I_3&=&\int_0^1 x^{{1\over 2}(m+1)}(1-x)^{{1\over 2}}{m\over\xi\sqrt{x}}J_m(\xi\sqrt{x})\nu^m_p(x)dx\nonumber\\
&=&{\sqrt{2}\Gamma(m+1)\Gamma(m+{3\over 2})\over(2p+m+{3\over 2})\Gamma(p+m+1)}m\left[{J_{2p+m+{1\over 2}}(\xi)\over \xi^{{3\over 2}}}+{J_{2p+m+{5\over 2}}(\xi)\over \xi^{{3\over 2}}}\right]\\
I_4&=&\int_0^1 x^{{1\over 2}(m+1)}(1-x)^{{1\over 2}}J_m^\prime(\xi\sqrt{x})\nu^m_p(x)dx\nonumber\\
&=&{\sqrt{2}\Gamma(m+1)\Gamma(p+{3\over 2})\over(2p+m+{3\over 2})\Gamma(p+m+1)}\nonumber\\
& &\times\left[(2p+m){J_{2p+m+{1\over 2}}(\xi)\over \xi^{{3\over 2}}}-(2p+m+3){J_{2p+m+{5\over 2}}(\xi)\over \xi^{{3\over 2}}}\right]
\end{eqnarray}
where we have used the relations
\begin{eqnarray}
I^{m-1,m}_{n+1,n}&=&\int_0^1 x^m(1-x)^{{1\over 2}}\nu^{m-1}_{n+1}(x)\nu^m_n(x)dx\nonumber\\
&=&-{\Gamma(m)\Gamma(m+1)\Gamma(n+2)\Gamma(n+{5\over 2})\Gamma(2n+m+{3\over 2})\over \Gamma(2n+m+{7\over 2})\Gamma(n+m+1)\Gamma(m+n+{3\over 2})}\\
I^{m-1,m}_{n,n}&=&\int_0^1 x^m(1-x)^{{1\over 2}}\nu^{m-1}_n(x)\nu^m_n(x)dx\nonumber \\
&=&{\Gamma(m)\Gamma(m+1)\Gamma(n+1)\Gamma(n+{3\over 2})\Gamma(2n+m+{1\over 2})\over \Gamma(2n+m+{5\over 2})\Gamma(n+m)\Gamma(m+n+{1\over 2})}\\
I^{m+1,m}_{n-1,n}&=&\int_0^1 x^{m+1}(1-x)^{{1\over 2}}\nu^{m+1}_{n-1}(x)\nu^m_n(x)dx\nonumber\\
&=&-{\Gamma(m+1)\Gamma(m+2n)\Gamma(n+1)\Gamma(n+{3\over 2})\Gamma(m+2n+{1\over 2})\over \Gamma(2n+m+{5\over 2})\Gamma(n+m+1)\Gamma(m+n+{3\over 2})}\\
I^{m+1,m}_{n,n}&=&\int_0^1 x^{m+1}(1-x)^{{1\over 2}}\nu^{m+1}_n(x)\nu^m_n(x)dx\nonumber\\
&=&{\Gamma(m+1)\Gamma(m+2)\Gamma(n+1)\Gamma(n+{3\over 2})\Gamma(2n+m+{3\over 2})\over \Gamma(2n+m+{7\over 2})\Gamma(n+m+1)\Gamma(m+n+{3\over 2})}
\end{eqnarray}
\section{Projection of $J_{m\pm1}(\xi\sqrt{x})$ into $u^m_n(x)$}
\begin{eqnarray}
K_1&=&\int_0^1 x^{{1\over 2}(m+1)}(1-x)^{-{1\over 2}}J_{m-1}(\xi\sqrt{x})u^m_p(x)dx\nonumber\\
&=&\sum_{q=0}^\infty{\sqrt{2}(2q+m-{1\over 2})\Gamma(q+m-{1\over 2})\over\Gamma(q+1)\Gamma(m)}{J_{2q+m-{1\over 2}}(\xi)\over \xi^{{1\over 2}}}K^{m-1,m}_{q,p}\nonumber\\
&=&{\sqrt{2}\Gamma(m+1)\Gamma(p+{1\over 2})\over (2p+m+{1\over 2})\Gamma(p+m)}{J_{2p+m-{1\over 2}}(\xi)\over \xi^{{3\over 2}}}-{\sqrt{2}\Gamma(m+1)\Gamma(p+{3\over 2})\over (2p+m+{1\over 2})\Gamma(p+m+1)}{J_{2p+m+{3\over 2}}(\xi)\over \xi^{{1\over 2}}}\nonumber\\
&=&{\sqrt{2}\Gamma(m+1)\Gamma(p+{1\over 2})\over (2p+m+{1\over 2})\Gamma(p+m+1)}\left[(p+m){J_{2p+m-{1\over 2}}(\xi)\over \xi^{{1\over 2}}}-\Bigl(p+{1\over 2}\Bigr){J_{2p+m+{3\over 2}}(\xi)\over \xi^{{1\over 2}}}\right]\\
K_2&=&\int_0^1 x^{{1\over 2}(m+1)}(1-x)^{-{1\over 2}}J_{m+1}(\xi\sqrt{x})u^m_p(x)dx\nonumber\\
&=&\sum_{q=0}^\infty{\sqrt{2}(2p+m+{3\over 2})\Gamma(q+m+{3\over 2})\over\Gamma(q+1)\Gamma(m+2)}{J_{2q+m+{3\over 2}}(\xi)\over \xi^{{1\over 2}}}K^{m+1,m}_{q,p}\nonumber\\
&=&-{\sqrt{2}p\Gamma(m+1)\Gamma(p+{1\over 2})\over (2p+m+{1\over 2})\Gamma(p+m+1)}{J_{2p+m-{1\over 2}}(\xi)\over \xi^{{1\over 2}}}+{\sqrt{2}(p+m+{1\over 2})\Gamma(m+1)\Gamma(p+{1\over 2})\over(2p+m+{1\over 2})\Gamma(p+m+1)}{J_{2p+m+{3\over 2}}(\xi)\over \xi^{{1\over 2}}}\nonumber\\
&=&{\sqrt{2}p\Gamma(m+1)\Gamma(p+{1\over 2})\over(2p+m+{1\over 2})\Gamma(p+m+1)}\left[-p{J_{2p+m-{1\over 2}}(\xi)\over \xi^{{1\over 2}}}+\Bigl(p+m+{1\over 2}\Bigr){J_{2p+m+{3\over 2}}(\xi)\over \xi^{{1\over 2}}}\right]\\
K_3&=&\int_0^1 x^{{1\over2}(m+1)}(1-x)^{-{1\over 2}}{m\over\xi\sqrt{x}}J_m(\xi\sqrt{x})u^m_p(x)dx\nonumber\\
&=&{\Gamma(m+1)\Gamma(p+{3\over 2})\over\sqrt{2}(2p+m+{1\over 2})\Gamma(p+m+1)}m\left[{J_{2p+m-{1\over 2}}(\xi)\over\xi^{{1\over 2}}}+{J_{2p+m+{3\over 2}}(\xi)\over \xi^{{1\over 2}}}\right]\nonumber\\
&=&{\Gamma(m+1)\Gamma(p+{3\over 2})\over\sqrt{2}(2p+m+{1\over 2})\Gamma(p+m+1)2m\Bigl(m+2p+{1\over 2}\Bigr){J_{2p+m+{1\over 2}}(\xi)\over\xi^{{3\over 2}}}}\\
K_4&=&\int_0^1 x^{{1\over 2}(m+1)}(1-x)^{-{1\over 2}}J_m^\prime(\xi\sqrt{x})u^m_p(x)dx \nonumber\\
&=&{\Gamma(m+1)\Gamma(p+{1\over 2})\over\sqrt{2}(2p+m+{1\over 2})\Gamma(p+m+1)}\left[(2p+m){J_{2p+m-{1\over 2}}(\xi)\over\xi^{{1\over 2}}}-(2p+m+1){J_{2p+m+{3\over 2}}(\xi)\over\xi^{{1\over 2}}}\right]
\end{eqnarray}
where
\begin{eqnarray}
K^{m-1,m}_{n+1,n}&=&\int^1_0 x^m(1-x)^{-{1\over 2}}u^{m-1}_{n+1}(x)u^m_n(x)dx\nonumber\\
&=&-{\Gamma(m)\Gamma(m+1)\Gamma(n+2)\Gamma(n+{3\over 2})\Gamma(2n+m+{1\over 2})\over\Gamma(2n+m+{5\over 2})\Gamma(n+m+1)\Gamma(m+n+{1\over 2})}\\
K^{m-1,m}_{n,n}&=&\int^0_1 x^m(1-x)^{-{1\over 2}}u^{m-1}_n(x)u^m_n(x)dx\nonumber\\
&=&{\Gamma(m)\Gamma(m+1)\Gamma(n+1)\Gamma(n+{1\over 2})\Gamma(2n+m-{1\over 2})\over\Gamma(2n+m+{3\over 2})\Gamma(n+m)\Gamma(m+n-{1\over 2})}\\
K^{m+1,m}_{n-1,n}&=&\int^1_0 x^{m+1}(1-x)^{-{1\over 2}}u^{m+1}_{n-1}(x)u^m_n(x)dx\nonumber\\
&=&-{\Gamma(m+1)\Gamma(m+2)\Gamma(n+1)\Gamma(n+{1\over 2})\Gamma(2n+m-{1\over 2})\over\Gamma(2n+m+{3\over 2})\Gamma(n+m+1)\Gamma(m+n+{1\over 2})}\\
K_{n,n}^{m+1,m}&=&\int^1_0 x^{m+1}(1-x)^{-{1\over 2}}u^{m+1}_n(x)u^m_n(x)dx\nonumber\\
&=&{\Gamma(m+1)\Gamma(m+2)\Gamma(n+1)\Gamma(n+{1\over 2})\Gamma(m+2n+{1\over 2})\over\Gamma(2n+m+{5\over 2})\Gamma(n+m+1)\Gamma(m+n+{1\over 2})}
\end{eqnarray}

\end{document}